\documentclass[prd,tightenlines,floatfix,showpacs,preprintnumbers,nofootinbib,eqsecnum,superscriptaddress,twocolumn]{revtex4-1}

\usepackage{color}      
\usepackage[dvips]{graphicx}    
\usepackage{amsbsy}     
\usepackage{amsfonts}       
\usepackage{amsmath}        
\usepackage{amssymb}        
\usepackage{color}
\usepackage{wasysym}
\usepackage{multirow}
\usepackage{mciteplus}
\usepackage{psfrag}
\usepackage{slashed}
\usepackage{cancel}

\newcommand{\MO}{{\tt micrOMEGAs}}
\newcommand{\DS}{{\tt DarkSUSY}}
\newcommand{\CHep}{{\tt CalcHEP}}

\newcommand{\SPheno}{{\tt SPheno}}

\newcommand{\FeynArts}{{\tt FeynArts}}
\newcommand{\FORM}{{\tt Form}}

\newcommand{\FeynCalc}{{\tt FeynCalc}}

\newcommand{\GeV}{{\;\mathrm{GeV}}}

\newcommand{\dd}{{\rm d}}

\catcode`\@=11
\font\manfnt=manfnt
\def\Watchout{\@ifnextchar [{\W@tchout}{\W@tchout[1]}}
\def\W@tchout[#1]{{\manfnt\@tempcnta#1\relax%
  \@whilenum\@tempcnta>\z@\do{%
    \char"7F\hskip 0.3em\advance\@tempcnta\m@ne}}}
\let\foo\W@tchout
\def\dubious{\@ifnextchar[{\@dubious}{\@dubious[1]}}

\def\@dubious[#1]{%
  \setbox\@tempboxa\hbox{\@W@tchout#1}
  \@tempdima\wd\@tempboxa
  \list{}{\leftmargin\@tempdima}\item[\hbox to 0pt{\hss\@W@tchout#1}]}
\def\@W@tchout#1{\W@tchout[#1]}
\catcode`\@=12

\begin{document}
\preprint{DESY 12-205, LAPTH-051/12, LPSC-12-341, MS-TP-12-17, KA-TP-42-2012}

\title{Neutralino-stop co-annihilation into electroweak gauge and Higgs bosons at one loop}


\author{J.~Harz}
 \email{julia.harz@desy.de}
 \affiliation{
 	Deutsches Elektronen-Synchrotron (DESY), Notkestra{\ss}e 85, D-22607 Hamburg, Germany
  }

\author{B.~Herrmann}
 \email{herrmann@lapth.cnrs.fr}
 \affiliation{
	LAPTh, Universit\'e de Savoie\,/\,CNRS, 9 Chemin de Bellevue, B.P.\ 110, F-74941 Annecy-le-Vieux, France
  }

\author{M.~Klasen}
 \email{michael.klasen@uni-muenster.de}
 \affiliation{
	Institut f\"ur Theoretische Physik, Westf\"alische Wilhelms-Universit\"at M\"unster, Wilhelm-Klemm-Stra{\ss}e 9, D-48149 M\"unster, Germany
  }

\author{K.~Kova\v{r}\'ik}
 \email{kovarik@particle.uni-karlsruhe.de}
 \affiliation{
	Institute for Theoretical Physics, Karlsruhe Institute of Technology, D-76128 Karlsruhe, Germany
  }
	
\author{Q.~Le Boulc'h}
 \email{leboulch@lpsc.in2p3.fr}
 \affiliation{
	Laboratoire de Physique Subatomique et de Cosmologie, Universit\'e Joseph Fourier/CNRS-IN2P3/INPG, 53 Rue des Martyrs, F-38026 Grenoble, France
  }

\date{\today}

\begin{abstract}
We compute the full O($\alpha_s$) supersymmetric QCD corrections for neutralino-stop
co-annihilation into electroweak gauge and Higgs bosons in the Minimal Supersymmetric
Standard Model (MSSM). We show that these annihilation channels are phenomenologically
relevant within the so-called phenomenological MSSM, in particular in the light of the
observation of a Higgs-like particle with a mass of about 126 GeV at the LHC. We present
in detail our calculation, including the renormalization scheme, the infrared treatment,
and the kinematical subtleties to be addressed. Numerical results for the co-annihilation
cross sections and the predicted neutralino relic density are presented. We demonstrate
that the impact of including the corrections on the cosmologically preferred region of
parameter space is larger than the current experimental uncertainty from WMAP data.
\end{abstract}


\maketitle


\section{Introduction}
\label{Sec:Intro}
Many astrophysical observations over a wide range of length scales provide today
convincing evidence of a sizable Cold Dark Matter (CDM) component in the
Universe. The most recent measurements of the WMAP satellite in combination
with baryonic acoustic oscillation and supernova data \cite{WMAP7} constrain the
dark matter relic density to the very precise value of 
\begin{equation}
	\Omega_{\rm CDM}h^2 ~=~ 0.1126 ~\pm~ 0.0036
	\label{Eq:WMAP}
\end{equation}
at $1\sigma$ confidence level,
where $h$ denotes the present Hubble expansion rate $H_0$ in units of 100 km
s$^{-1}$\,Mpc$^{-1}$. The leading candidate for dark matter is a weakly
interacting massive particle (WIMP). Unfortunately, the Standard Model does not
contain any candidates that would be compatible with the properties of cold
dark matter, the neutrinos being too light.

In contrast, the Minimal Supersymmetric Standard Model (MSSM) with conserved
$R$-parity contains with the lightest neutralino, denoted $\tilde{\chi}^0_1$, a
stable WIMP. Since supersymmetry offers many other theoretical and
phenomenological advantages, the neutralino is by far the most studied dark
matter candidate. The time evolution of its number density $n_{\chi}$ is
described by the Boltzmann equation
\begin{equation}\label{Eq:Boltzmann}
 \frac{\dd n_{\chi}}{\dd t} ~=~ - 3 H n_{\chi} - \langle \sigma_{\rm ann}v \rangle
 \left[ n_{\chi}^2 - (n_{\chi}^{\rm eq})^2 \right] ,
\end{equation}
where the first term on the right-hand-side containing the Hubble parameter
$H$ stands for the dilution of dark matter due to the expansion of the
Universe. The remaining terms reduce (increase) the number of dark matter
particles by their annihilation (creation) in collisions with other particles.
The term $\langle \sigma_{\rm ann}v \rangle$ denotes the thermal average of the
annihilation cross section of the neutralino
\begin{eqnarray}
 &&\langle \sigma_{\rm ann}v \rangle = \frac{\int \sigma v\ e^{-E_1/T}e^{-E_2/T}
 \dd^3p_1 \dd^3p_2}{\int e^{-E_1/T}e^{-E_2/T} \dd^3p_1 \dd^3p_2}\\ \nonumber
 &&= \frac{1}{8m^4 T K_2^2\big(m/T\big)}\int_{4m^2}^\infty \sigma \big(s-4m^2\big)
 \sqrt{s}\,K_1\big(\sqrt{s}/T\big)\, \dd s\,,
\end{eqnarray}
where $K_i$ are the modified Bessel functions of the second kind of order $i$ 
(for details see Ref.\ \cite{GondoloGelmini}).

Here we will consider the case when heavier, unstable supersymmetric particles
survive in the Universe for sufficient time to affect the relic density of the
dark matter particle. In this case, Eq.~(\ref{Eq:Boltzmann}) has to be modified
to account for the interactions between all particles and solve a system of
Boltzmann equations for number densities $n_i$ for each surviving particle
species,
\begin{equation}
 \frac{\dd n_{i}}{\dd t} ~=~ - 3 H n_{i} - \langle \sigma_{ij}v_{ij} \rangle
 \left( n_i n_j - n_i^{\rm eq}n_j^{\rm eq} \right) ,
\end{equation}
where $\sigma_{ij} \equiv \sigma (\chi_i \chi_j \rightarrow X)$ is the cross
section of the annihilation of particle $i$ with particle $j$, and $v_{ij}$ is
their relative velocity. As all heavier particles eventually decay into the
dark matter particle, the relevant quantity is the total number density
$n_\chi=\sum_i n_i$, and we can reformulate the problem into solving a single
Boltzmann equation similar to Eq.~(\ref{Eq:Boltzmann}) for the total number
density with the annihilation cross section $\langle \sigma_{\rm ann}v \rangle$
replaced by an effective cross section $\langle \sigma_{\rm eff}v \rangle$.
More precisely, this cross section is given by
\begin{equation}
 \langle \sigma_{\rm eff}v \rangle ~=~ \sum_{i,j} \sigma_{ij}v_{ij}
 \frac{n^{\rm eq}_i}{n_{\chi}^{\rm eq}} \frac{n^{\rm eq}_j}{n_{\chi}^{\rm eq}}\,,
 \label{Eq:sigmaeff}
\end{equation}
where the sum runs over all MSSM particles $i$ and $j$ (for a detailed
discussion see Refs.\ \cite{GriestSeckel,EdsjoGondolo}). The ratio of their
respective number density in thermal equilibrium, $n_i^{\rm eq}$, and the
number density of the neutralino, $n_{\chi}^{\rm eq}$, at the temperature $T$ is
Boltzmann suppressed,
\begin{equation}
 \frac{n^{\rm eq}_i}{n_{\chi}^{\rm eq}} ~\sim~ \exp\left[ - \frac{m_i-m_{\chi}}{T}
 \right] , \label{Eq:BoltzmannSupression}
\end{equation}
so that only particles, whose masses are almost degenerate with the one of
lightest neutralino, can give sizable contributions. In the MSSM, typical
examples of relevant co-annihilations are those of the neutralino with the
lightest slepton or squark, or with another gaugino. Moreover, pair
annihilations of the next-to-lightest superpartner can be non-negligible.
Having solved the Boltzmann equation numerically, the relic density is
finally obtained through
\begin{equation}
 \label{Eq:OmegaMass}
 \Omega_{\chi}h^2 ~=~ \frac{m_{\chi} n_{\chi}}{\rho_{\rm crit}}\,,
\end{equation}
where $n_{\chi}$ is the current number density of the neutralino and
$\rho_{\rm crit}$ is the critical density of the Universe. Comparing the
predicted value obtained by solving the Boltzmann equation to the
observational limits in Eq.~(\ref{Eq:WMAP}) allows one to identify
cosmologically (dis-)favored regions of the MSSM parameter space and thus
to obtain important information that is complementary to collider searches
and precision measurements.

The procedure described above is unfortunately subject to several
uncertainties. The first source of uncertainty lies in the extraction of
the relic density of CDM from cosmological data as given in
Eq.~(\ref{Eq:WMAP}). The extraction is based on a simple cosmological model,
the $\Lambda$CDM model, which uses a minimal set of six parameters to fit
the available cosmological data and bases its conclusions on the Standard
Model of cosmology \cite{WMAP7}. It has been shown that changing either the
number of free parameters of the model used to fit the cosmological data
\cite{Hamann} or modifying the assumptions contained in the Standard Model
of cosmology (e.g., altering the expansion rate in the primordial Universe
or later, but still before Big Bang Nucleosynthesis \cite{Arbey}), may
change the extracted central value of $\Omega_{\rm CDM}h^2$ along with the
confidence levels. 

The second source of uncertainty in identifying (dis\nobreakdash-) favored regions of
the MSSM parameter space is connected to the calculation of the essential
parameters such as masses and couplings of supersymmetric particles. As
the relic density is very sensitive to the mass of the neutralino (see
Eq.~(\ref{Eq:OmegaMass})), any uncertainty in the calculation of its mass
directly translates into an uncertainty on the calculated relic density.
Moreover, the relic density also strongly depends on the (co-)annihilation
cross section, which in turn crucially depends on the masses of the
remaining particles and their couplings to the neutralino. In the MSSM,
the mass and the couplings of the neutralino as well as any other relevant
couplings and masses are typically obtained using a dedicated spectrum
calculator (see, e.g., Ref.\ \cite{SPheno}), which evolves all parameters down
from a grand unification scale and calculates the masses and couplings for
all particles at the weak scale. Different treatments of the radiative
corrections for masses and couplings as well as different implementations
of the renormalization group equations in various MSSM spectrum
calculations can lead to differences in the predictions for the relic
density and thus in the preferred/excluded regions of the MSSM parameter
space (for details see, e.g., Ref.\ \cite{Belanger}).

The uncertainty which we will address in this paper does not fall into
either of the above mentioned categories, but concerns the precision,
with which the (co\nobreakdash-) annihilation cross sections in Eq.\ (\ref{Eq:sigmaeff})
are computed. The cross sections in public dark matter tools such as {\DS}
\cite{DarkSusy} or {\MO} \cite{micrOMEGAs2007} are implemented using only
an effective tree-level calculation. It is, however, well known that
higher-order corrections, particularly those involving the strong coupling
constant, can have a sizable impact on such processes.
The impact of next-to-leading order corrections to neutralino annihilation
on the neutralino relic density has been discussed in several previous
analyses. SUSY-QCD corrections to neutralino pair annihilation into
quark-antiquark pairs have been studied in Refs.\ \cite{DMNLO_AFunnel,
DMNLO_mSUGRA, DMNLO_NUHM}, while the corresponding electroweak corrections
have been evaluated in Refs.\ \cite{Sloops2007, Sloops2009, Sloops2010}.
The authors of Refs.\ \cite{Sloops2009, Sloops2010} have also discussed the
case of co-annihilation of a neutralino with another gaugino. Further
studies rely on effective coupling approaches in order to capture certain
classes of corrections to neutralino pair annihilation and co-annihilation
with a tau slepton \cite{Sloops2011,EffCouplings}. All these analyses show that
radiative corrections are not negligible in the context of relic density
calculations, the impact of the corrections being larger than the
experimental uncertainty from WMAP in many regions of parameter space.
With the Planck satellite data providing more precise cosmological
measurements in the very near future, it becomes even more pressing that
theoretical predictions match the experimental precision.
\begin{figure*}[t]
 \includegraphics[scale=0.7]{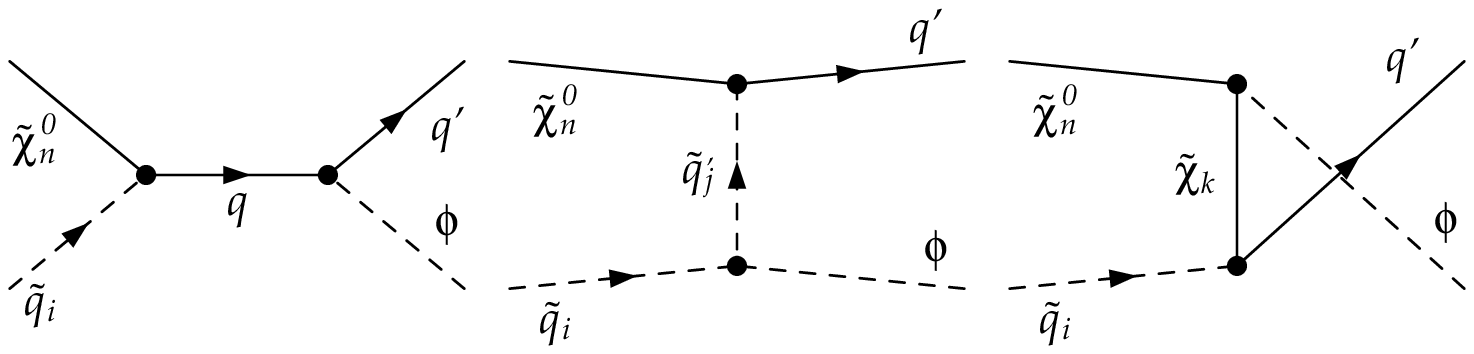}
 \includegraphics[scale=0.7]{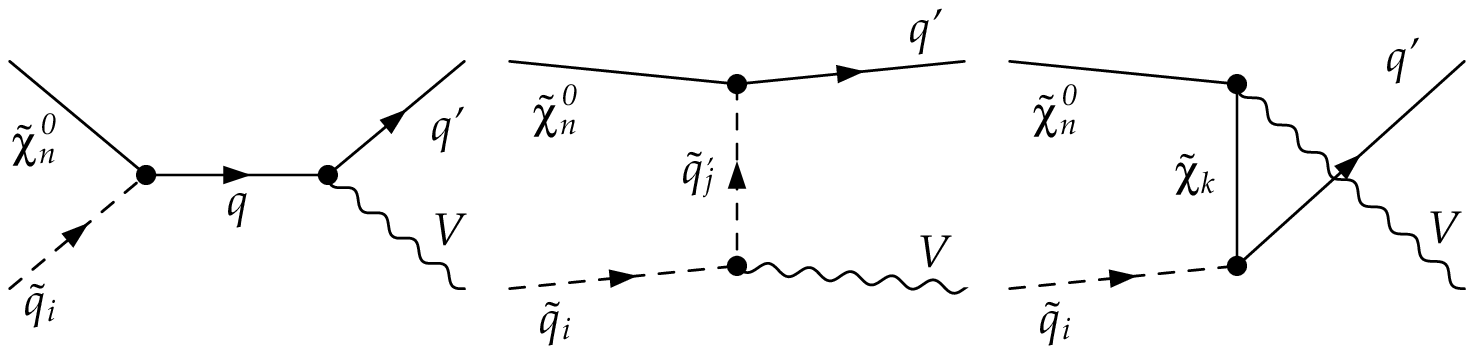}
 \caption{Leading-order Feynman diagrams for neutralino-squark
 co-annihilation into a quark and a Higgs boson ($\phi=h^0,H^0,A^0,H^{\pm}$)
 or an electroweak gauge boson ($V=\gamma,Z^0,W^{\pm}$). The $u$-channel is absent for a photon in the final state.}
 \label{Fig:TreeDiagrams}
\end{figure*}

The important case of SUSY-QCD corrections to co-annihilation of a
neutralino with a scalar top has so far only been considered in Ref.\
\cite{Freitas2007}. This study concerns the very specific cases of
co-annihilation of a bino-like neutralino with a right-handed stop into
a top quark and a gluon as well as into a bottom quark and a $W$-boson.
However, depending on the considered region of parameter space, many
other final states, including those with other electroweak gauge and Higgs
bosons, can become dominant. Moreover, in realistic supersymmetric
scenarios, helicity mixing in the stop sector is usually non-negligible,
as is the mixing of bino, wino, and higgsino components in the lightest
neutralino, which strongly influences its couplings and preferred
(co-)annihilation channels. 
Therefore, we extend in this paper the analysis of QCD and SUSY-QCD
corrections to co-annihilation of a neutralino with a stop by computing
the general case of neutralino-stop co-annihilation into a quark and a
Higgs or an electroweak vector boson. The paper is
organized as follows: In Sec.\ \ref{Sec:Pheno}, we first discuss the
phenomenology of neutralino-stop co-annihilation in the MSSM. We then
describe in detail the calculation of the radiative corrections to the
relevant processes in Sec.\ \ref{Sec:Calculation}. Numerical results
for annihilation cross sections and dark matter relic densities in
typical MSSM benchmark scenarios are presented in Sec.\
\ref{Sec:Results}, and our conclusions are given in Sec.\
\ref{Sec:Conclusions}.

\section{Phenomenology of neutralino-stop co-annihilation \label{Sec:Pheno}}

As discussed in Sec.\ \ref{Sec:Intro}, the co-annihilation of the
next-to-lightest supersymmetric particle (NLSP) with the lightest
neutralino can in certain regions of the MSSM parameter space become
dominant and lead to a relic density that is compatible with the
observational limit of Eq.\ (\ref{Eq:WMAP}). A particularly important
example of such an NLSP is the scalar top, whose chirality eigenstates
can mix significantly, e.g.\ when the trilinear coupling $A_t$ becomes
large, and which can then have a lower mass eigenstate that is almost
mass-degenerate with the lightest neutralino \cite{StopCoann2, StopCoann1}.

There is ample motivation for a light scalar top. First, a light
stop is a necessary ingredient to achieve electroweak baryogenesis in
the MSSM \cite{EWBG}. Second, ``natural'' SUSY models
\cite{naturalSUSY1, naturalSUSY2} require a light third generation of
sfermions in order to reduce fine-tuning and stay compatible with
experimental constraints at the same time. This is due to the fact that
the mass degeneracy between the lightest neutralino and NLSP weakens
the LHC exclusion potential on the third-generation squark masses, since
this degeneracy results in events with soft jets
\cite{StopCoannLHC1, StopCoannLHC2}. Third, interpreting the new boson
with a mass of about 126 GeV observed recently at the LHC
\cite{ATLAS2012, CMS2012, ATLAS2012update} as a light CP-even Higgs boson ($h^0$)
implies within the MSSM a particular choice of parameters in the
stop and sbottom sector \cite{Arbey2012}. The reason is that in the MSSM
the lightest Higgs boson mass receives a large contribution from a loop
containing scalar tops. The leading contribution to the mass coming from 
this loop together with the tree-level contribution can be expressed as \cite{Haber1996, Badziak2012}
\begin{widetext}
\begin{equation}
 m_{h^0}^2 ~=~ m_Z^2 \cos^2 2\beta + \frac{3 g^2 m_t^4}{8 \pi^2 m_W^2} 
 \left[ \log\frac{M_{\rm SUSY}^2}{m_t^2} + \frac{X^2_t}{M^2_{\rm SUSY}}
 \left( 1 - \frac{X^2_t}{12 \,M^2_{\rm SUSY}} \right) \right]\,,
 \label{Eq:HiggsMass}
\end{equation}
\end{widetext}
with $X_t = A_t - \mu / \tan\beta$ and $M_{\rm SUSY} = \sqrt{m_{\tilde{t}_1}
m_{\tilde{t}_2}}$. The maximal contribution from stop mixing is then
obtained for $|X_t| \sim \sqrt{6} M_{\rm SUSY}$, which favors a sizable
trilinear coupling $A_t$ and consequently a rather light stop. 

At tree level, the co-annihilation of a neutralino and a stop into final
states containing a quark and an electroweak gauge or Higgs boson is
mediated either by an $s$-channel quark, a $t$-channel squark, or a
$u$-channel neutralino or chargino exchange. The corresponding Feynman
diagrams are depicted in Fig.~\ref{Fig:TreeDiagrams}. These processes
compete with all other possible (co-)annihilation channels of the
lightest neutralino and in certain cases also with stop pair annihilation. 
\begin{figure*}[t]
 \begin{center}
 \includegraphics[scale=0.85]{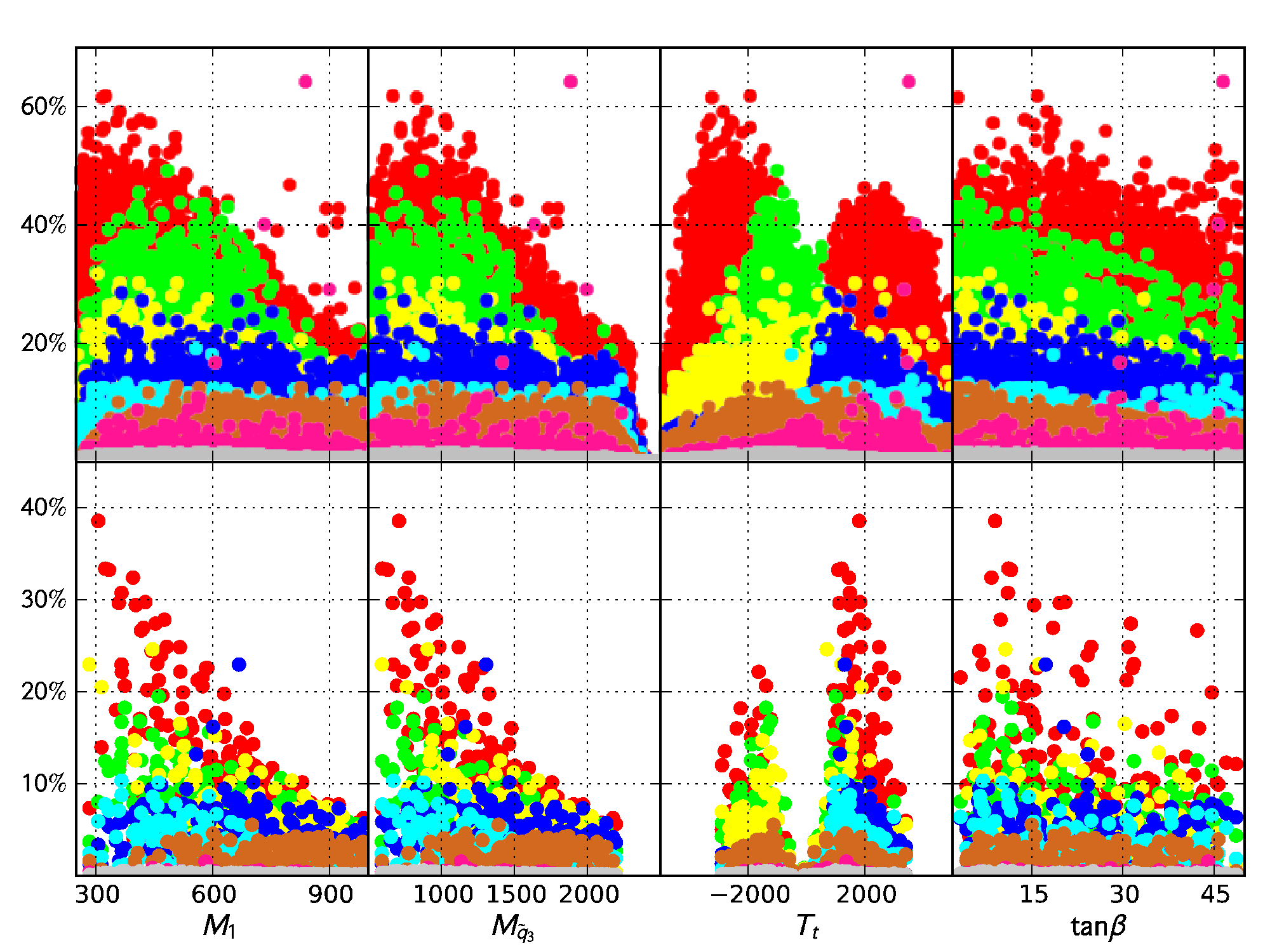}
 \vspace*{-8mm}
 \end{center}
 \caption{Relative contributions of the neutralino-stop co-annihilation
 channels for the generated parameter points as a function of the input
 parameters $M_1$, $M_{\tilde{q}_3}$, $T_t$, and $\tan\beta$ before (top) and
 after (bottom) applying the selection cuts of Eq.\ (\ref{Eq:SelectionCuts}).
 Shown are the contributions from $th^0$ (red), $tg$ (green), $tZ^0$ (blue),
 $tH^0$ (yellow), $bW^+$ (cyan), $tA^0$ (brown), $bH^+$ (pink), and
 $t\gamma$ (gray) final states. The parameters $M_1$, $M_{\tilde{q}_3}$, and
 $T_t$ are given in GeV.}
 \label{Fig:RandomScan}
\end{figure*}

In order to quantify the relative importance of the processes in 
Fig.~\ref{Fig:TreeDiagrams}, we have performed a random scan in the
phenomenological MSSM. In the following we describe the settings and discuss in detail the results of our scan.
According to the SPA convention \cite{SPA2005} the soft-breaking parameters are defined
at the scale $Q=1$ TeV. We have made a few simplifying assumptions, which bring the number of
parameters down to eight. In the squark sector, we use a common mass
parameter $M_{\tilde{q}_{1,2}}$ for the squarks of the first and second
generation, leaving the common mass parameter $M_{\tilde{q}_3}$ for the
left- and right-handed squarks of the third generation independent.
In contrast, the slepton sector is characterized by a single mass
parameter $M_{\tilde{\ell}}$ for all three generations. All trilinear
couplings are set to zero except for the $A_t$ in the stop sector, which
enters our calculations through the relation $T_t = Y_t A_t$ with 
the top Yukawa coupling $Y_t$. All gaugino masses
are defined through the bino mass parameter $M_1$. The wino and gluino
masses are then fixed by the relation $2 M_1 = M_2 = M_3/3$, which is
deduced from gaugino mass unification at the GUT scale.
Finally, the Higgs sector is specified by
the pole mass $m_A$ of the pseudoscalar Higgs boson, the higgsino mass
parameter $\mu$, as well as the ratio $\tan\beta$ of the two vacuum
expectation values of the Higgs doublets. 
In order to explore the parameter space, we have randomly generated
$1.2$ million parameter points within the following ranges for the eight
input parameters:
\begin{eqnarray}
 500~{\rm GeV} \leq M_{\tilde{q}_{1,2}} &\leq&~ 4000~{\rm GeV}, \nonumber\\
 100~{\rm GeV} \leq M_{\tilde{q}_3}     &\leq&~ 2500~{\rm GeV} , \nonumber\\
 500~{\rm GeV} \leq M_{\tilde{\ell}}   ~&\leq&~ 4000~{\rm GeV}, \nonumber \\
                    |T_t|              ~&\leq&~ 5000~{\rm GeV} ,  \nonumber\\
 200~{\rm GeV} \leq M_1                ~&\leq&~ 1000~{\rm GeV}, \\
 100~{\rm GeV} \leq m_A                ~&\leq&~ 2000~{\rm GeV} ,  \nonumber\\
                            |\mu|      ~&\leq&~ 3000~{\rm GeV} , \nonumber \\
 2             \leq        \tan\beta    &\leq&  50 . \nonumber
\end{eqnarray}
\begin{table*}[t]
 \begin{tabular}{|c|cccccccc|cccc|}
 \hline
 & ~~~ $M_1$ ~~~& ~~~$M_{\tilde{q}_{1,2}}$~~~ & ~~~$M_{\tilde{q}_3}$~~~ & ~~~$M_{\tilde{\ell}}$~~~ & ~~~$T_t$~~~ & ~~~$m_{A}$~~~ &
 ~~~$\mu$~~~ & ~~~$\tan\beta$~~~ & ~~~$m_{\tilde{\chi}^0_1}$~~~ & ~~~$m_{\tilde{t}_1}$~~~ & ~~~$m_{h^0}$~~~ & ~~~$m_{H^0}$~~~  \\
 \hline
 I & $306.9$ & $2037.7$ & $709.7$ & $1499.3$ & $1806.5$ & $1495.6$ & $2616.1$ & $9.0$ & 307.1 & 350.0 & 124.43 & 1530.72 \\
 II & $470.6$ & $1261.2$ & $905.3$ & $1963.2$ & $1514.8$ & $1343.1$ & $725.9$ & $18.3$ & 467.3 & 509.4 & 124.06 & 1342.77 \\
 III & $314.4$ & $2870.5$ & $763.6$ & $2417.7$ & $1877.5$ & $386.0$ & $2301.5$ & $10.3$ & 316.5 & 371.9 & 123.43 & 367.45 \\
 \hline
 \end{tabular}
 \caption{Three characteristic scenarios chosen in the pMSSM, which will be considered
 in this study. Given are the input parameters as described in the text, the lightest
 neutralino mass $m_{\tilde{\chi}_1^0}$, the lightest stop mass $m_{\tilde{t}_1}$, and the
 masses of the light and heavy CP-even Higgs bosons $m_{h^0}$ and $m_{H^0}$. All values
 except for $\tan\beta$ are given in GeV.}
 \label{Tab:Scenarios}
\end{table*}
\begin{table*}[t]
 \begin{tabular}{|c|c|c|c|c|c|c|}
 \hline
  & ~~$\Omega_{\chi}h^2$~~ & ~~ $\tilde{\chi}^0_1 \tilde{t}_1 \rightarrow t h^0$ ~~ & ~~ $\tilde{\chi}^0_1 \tilde{t}_1 \rightarrow t H^0$ ~~ & ~~ $\tilde{\chi}^0_1 \tilde{t}_1 \rightarrow t Z^0$ ~~ & ~~ $\tilde{\chi}^0_1 \tilde{t}_1 \rightarrow b W^+$ ~~ & ~~ Sum ~~ \\
 \hline
 I   & 0.114 & 38.5\% & --     & 3.4\%  & 5.9\% & 47.8\% \\
 II  & 0.116 & 24.6\% & --     & 10.7\% & 3.4\% & 38.7\% \\
 III & 0.111 & 14.2\% & 20.7\% & 1.2\%  & 2.1\% & 38.2\% \\
 \hline	
 \end{tabular}
 \caption{Neutralino relic density and relative contributions of neutralino-stop
 co-annihilation into a quark and a Higgs or electroweak gauge boson for the
 characteristic scenarios of Tab.\ \ref{Tab:Scenarios}. The last column gives the
 sum of the listed contributions.}
 \label{Tab:Channels}
\end{table*}
For each set of parameters, the physical mass spectrum and the
related mixing matrices have been obtained using {\SPheno} \cite{SPheno} (version {\tt 3.2.1}).
The neutralino relic density $\Omega_{\chi}h^2$ as well as the contributions
from the individual (co-)annihilation channels have been computed using
{\MO} (version {\tt 2.4.1}).
For the numerical values of the Standard Model parameters we refer
the reader to Ref.\ \cite{PDG2012}.
For a substantial number of these scenarios, co-annihilation of the lightest
neutralino with a scalar top plays an important role. This can be seen in the
upper part of Fig.~\ref{Fig:RandomScan}, where we show the relative
contributions of the different final states channels to the total
(co-)annihilation cross section as a function on the phenomenologically most
relevant input parameters. 

Experimentally viable scenarios have to satisfy a number of additional
constraints. We therefore impose the following cuts on the neutralino relic
density, the mass of the lightest Higgs boson, and the inclusive branching
ratio of the most sensitive $B$-meson decay, $b\to s\gamma$:
\begin{gather}\nonumber
 0.0946 \leq~  \Omega_{\chi}h^2 ~\leq 0.1306 , \\ \label{Eq:SelectionCuts}
 120\,{\GeV} \leq~ m_{h^0} ~\leq 130\,{\GeV} , \\ \nonumber
 2.77 \cdot 10^{-4} \leq~ {\rm BR}(b\to s\gamma) ~\leq 4.33\cdot 10^{-4} .
\end{gather}
The first cut selects the points which match the observed relic density
of Eq.\ (\ref{Eq:WMAP}) within a $5\sigma$ confidence interval. The second limit
corresponds to a very conservative mass range for the new boson observed at
the LHC \cite{ATLAS2012, CMS2012}. Note that the theoretical uncertainty on
the calculation of the lightest Higgs boson mass within {\tt SPheno} is
estimated to be about 3 GeV \cite{SPheno}. Finally, the limits on the
branching ratio of $b\to s\gamma$ correspond to a $3\sigma$ interval around
the observed value of ${\rm BR}(b\to s\gamma)=(3.55 \pm 0.26)\cdot 10^{-4}$
\cite{HFAG}. 
The points selected in this way are depicted in the lower part of 
Fig.~\ref{Fig:RandomScan}, where we show again the relative contribution of the
different neutralino-stop co-annihilation channels. Applying the
experimental cuts described above reduces the density of the points, but
does not significantly change the shape of the distributions. As can be seen, the
statistically most important final state is a top quark together with a
light Higgs boson, followed by top quark and a gluon, a
heavy CP-even Higgs boson, or a $Z$-boson. Comparable in size to the latter channel is the
co-annihilation into a bottom quark and a $W$-boson, whereas final states
including a pseudoscalar Higgs boson, a charged Higgs boson, or a photon are less important.

One viable option how to satisfy the relic density bound and respect 
current exclusion limits from colliders is that the lightest neutralino 
and the lightest scalar top are almost mass degenerate. This is 
reflected in the left and left-center columns of Fig.~\ref{Fig:RandomScan}
where we can observe a strikingly similar dependence of fraction of 
co-annihilation processes on the gaugino mass parameter $M_1$ and 
the third-generation squark mass parameter $M_{\tilde{q}_3}$, which are 
largely responsible for the masses of neutralinos and squarks of the third 
generation. For large values of both parameters, co-annihilations cease to
to be important and annihilations of stops take their place as the dominating
contribution of the total cross section.
\begin{figure*}[t]
 \includegraphics[scale=0.42]{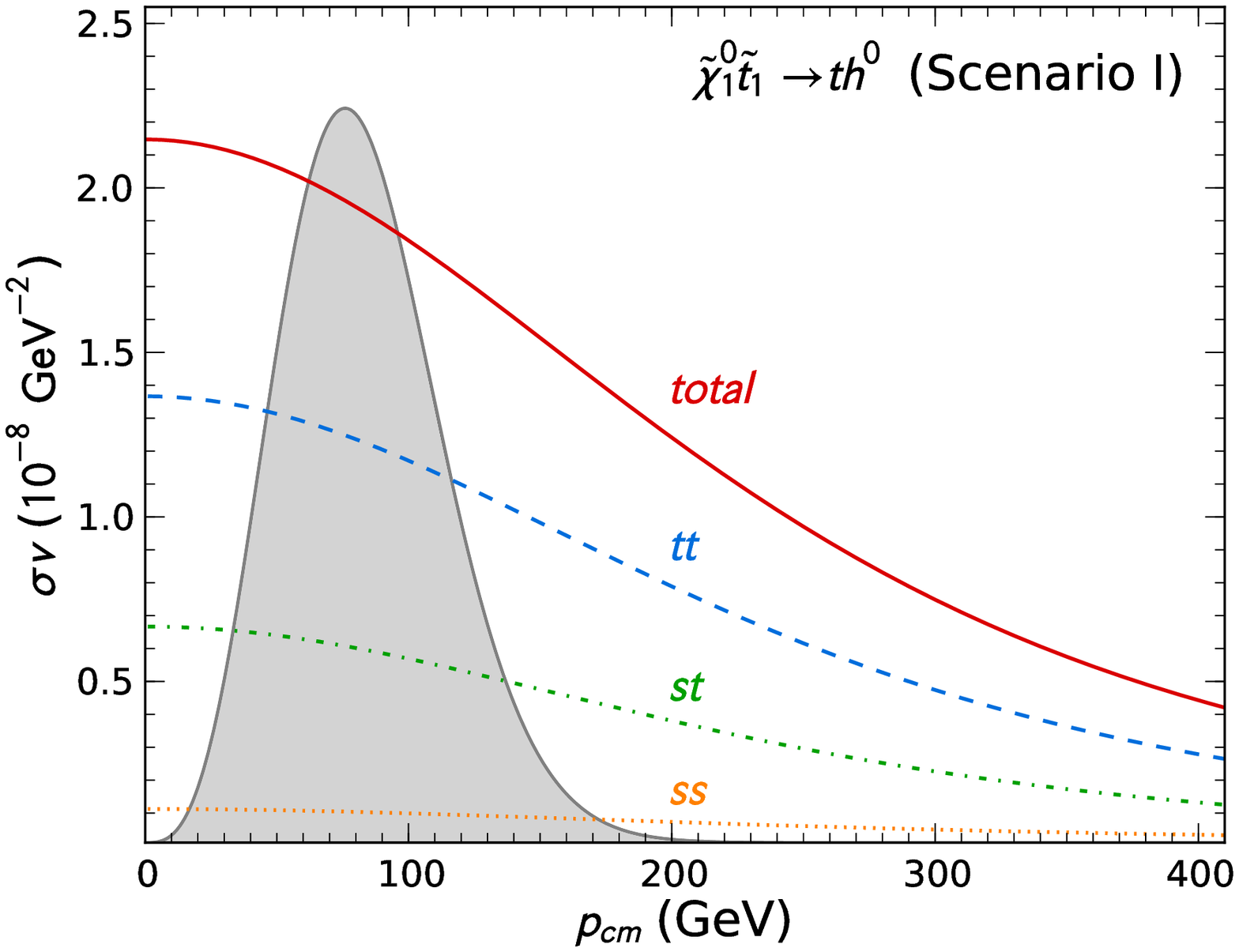}
 \includegraphics[scale=0.42]{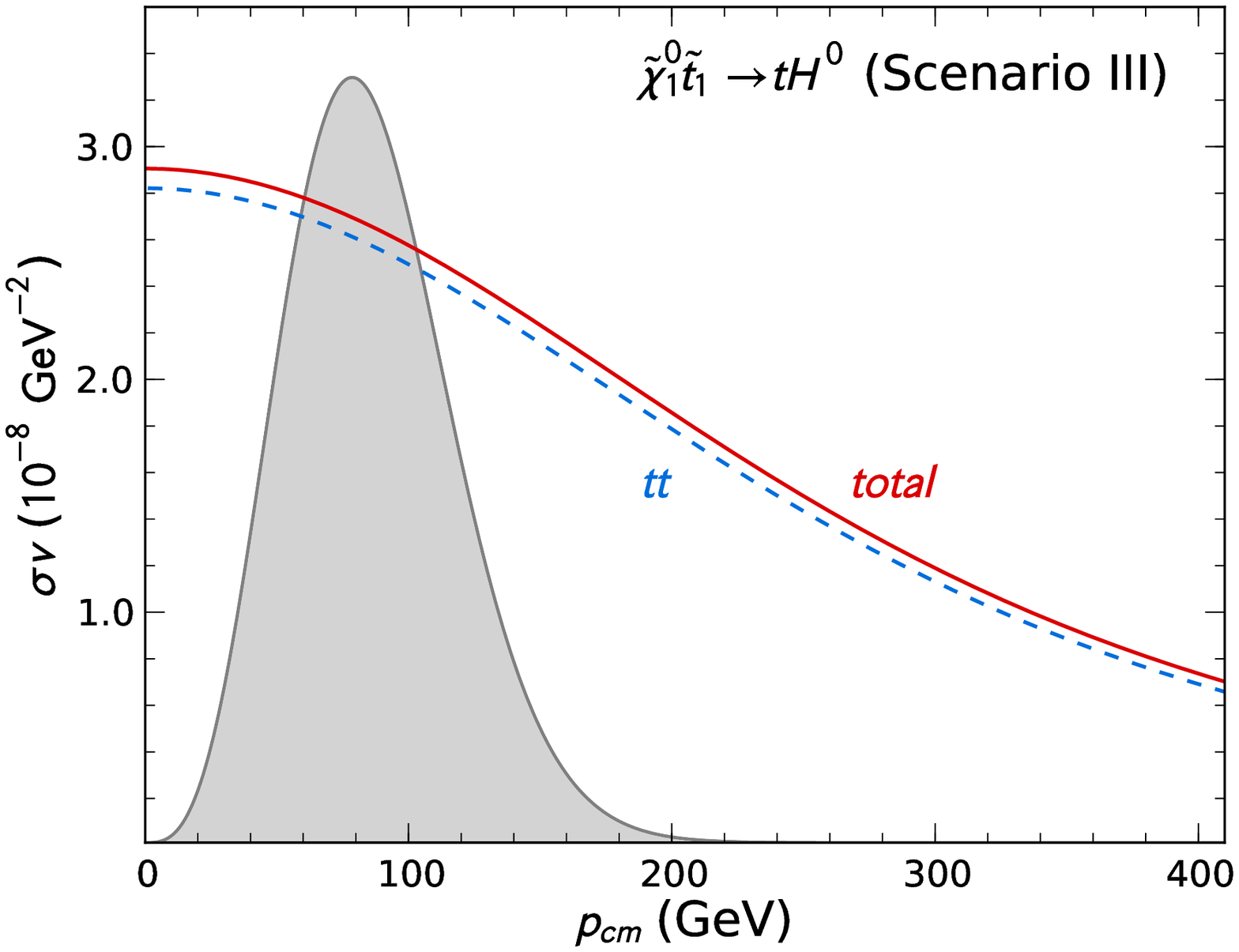} \\
 \includegraphics[scale=0.42]{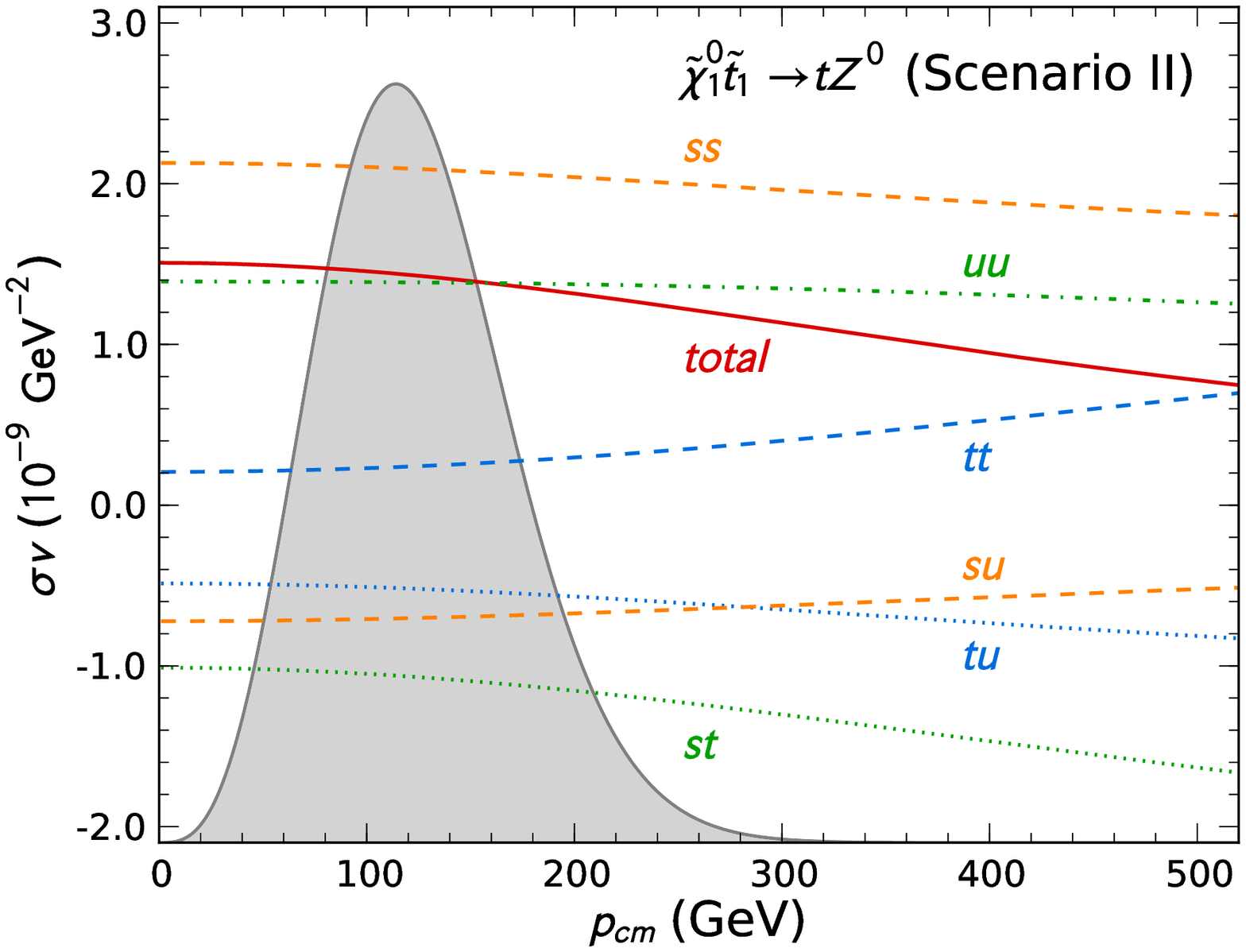}
 \includegraphics[scale=0.42]{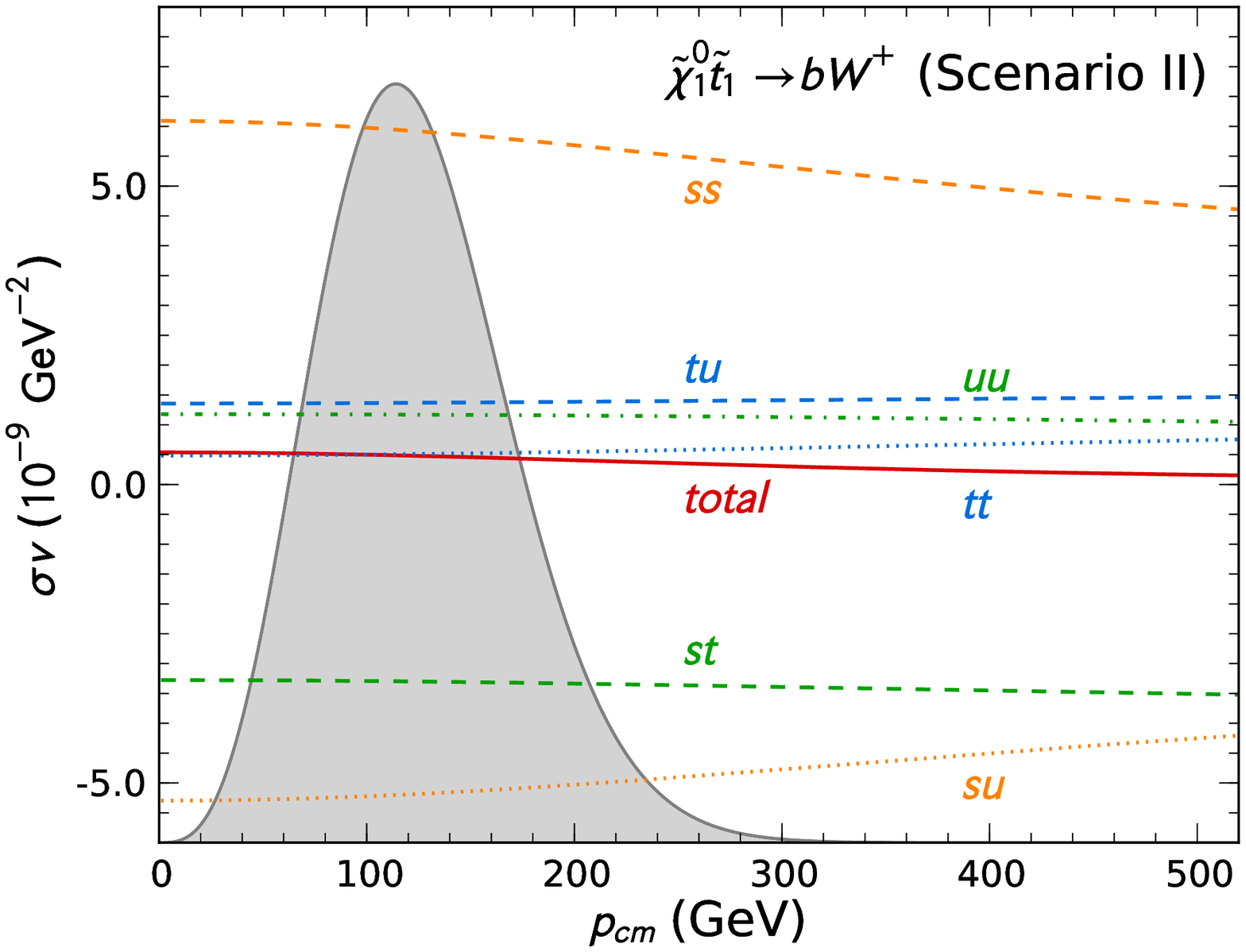}
 \caption{Contribution of the different diagrams ($s$-, $t$-, and $u$-channel) depicted in Fig.~\ref{Fig:TreeDiagrams}. For the studied scenarios of Tab.\ \ref{Tab:Scenarios} we show for selected co-annihilation channels the tree-level cross section as well as the contribution of the different squared diagrams ($ss$, $tt$, $uu$) and the interference terms ($st$, $su$, $tu$).}
 \label{Fig:TreeChannels}
\end{figure*}

In the right-center panel of Fig.~\ref{Fig:RandomScan} one can notice
the interesting feature that after applying the cut on the lightest
Higgs boson mass, large values of $T_t$ are preferred and the initially rather 
important percentage of top-gluon final states is reduced.  
This is driven by the fact that in contrast to the top-gluon final state, 
the Higgs boson mass prefers a sizable trilinear coupling.
Moreover, positive values for $T_t$ are slightly preferred, since they
allow a better maximization of the Higgs boson mass \cite{Arbey2012}.
This is well visible in the center-right column of Fig.~\ref{Fig:RandomScan},
where after applying the cuts two distinct ranges for the trilinear coupling
parameter $T_t$ can be observed. These large values also enhance the
Higgs-squark-squark coupling, which is present in the $t$-channel of the $th^0$ final
state. Accordingly, this changes the relative importance of the squark exchange
with respect to the two other diagrams (quark or neutralino exchange, see 
Fig.~\ref{Fig:TreeDiagrams}). The $t$-channel enhancement also leads to an almost
universal dominance of co-annihilation into Higgs final states in the scenarios
considered here. In other words, the same mechanism which drives the mass of
the lightest Higgs boson to the observed value through important stop-loop
contributions is responsible for the increase of neutralino-stop co-annihilation
into the lightest Higgs boson together with a top quark. 

A similar connection between parameters that we mentioned above for $M_1$ and $M_{\tilde{q}_3}$ 
is found for $T_t$ and the Higgs parameter $\mu$. After the cuts, large and positive values of $\mu$
are preferred, which also enhances the Higgs-sfermion-sfermion coupling 
mainly for the heavy CP-even Higgs with the same consequences as discussed above for large
values of $T_t$.

The dependence on $\tan\beta$, on the other hand, is generally less pronounced.
For co-annihilation (mostly into top quarks), lower values of $\tan\beta$
are slightly preferred, since $b\bar{b}$ final states become more important
for $\tan\beta\gtrsim 40$ \cite{DMNLO_AFunnel}. 
The influence of the remaining input parameters, such as those related to first 
and second generation squarks, sbottoms, and sleptons, as well as the higgsinos, 
is less important in this context. Therefore the corresponding dependencies are not displayed in 
Fig.~\ref{Fig:RandomScan}.

For our numerical analysis, we have selected three characteristic scenarios, which we introduce and discuss in the following.
They are listed in Tab.\ \ref{Tab:Scenarios} and have been chosen in such a way
that they represent qualitatively different scenarios (note, e.g., the differences
in $M_{\tilde{q}_{1,2}}$, $m_A$, and $\mu$) and that they lead to different dominant
co-annihilation final states. As expected from Eq.\ (\ref{Eq:HiggsMass}), all
three scenarios feature rather important trilinear coupling parameters
$T_t \sim 1500 - 1800$ GeV. The selected values of $\tan\beta$ are moderate, so
that neutralino pair annihilation into bottom quarks is not important here. First
and second generation squarks and sleptons are heavy compared to the stops in
accordance with current LHC exclusion limits \cite{asl,csl}.
Moreover, the mass difference of the lightest neutralino and the scalar top
is about 50 GeV in each scenario and thus sufficiently small. In Tab.\
\ref{Tab:Channels} we list the resulting values for the neutralino relic density,
together with the contributions from the neutralino-stop co-annihilation modes.
These will be crucial to estimate the impact of our calculations on the final
relic density.

Scenario I is characterized by the dominant co-annihilation into a top quark
and a light Higgs boson. Final states including a top quark and a $Z$-boson
as well as a bottom quark and a $W$-boson contribute as well, but to a lesser
extent. In total, neutralino-stop co-annihilation with electroweak gauge and
Higgs bosons final states accounts for almost half of the annihilation cross section
at this example point. In order to understand which diagrams of Fig.~\ref{Fig:TreeDiagrams}
are most important in this context, we show in Fig.~\ref{Fig:TreeChannels} the total
tree-level cross sections of neutralino-stop co-annihilation into the dominant final
states for each characteristic scenario, together with the individual contributions
of the different squared diagrams and interference terms. For the reasons
discussed above, the exchange of a scalar top in the $t$-channel is the
dominant mode at example point I, followed by its interference with the
exchange of a top quark in the $s$-channel (upper left plot). The squared
$s$-channel is rather small, and all other channels are even negligible in
this parameter configuration, so that they are not shown in Fig.~\ref{Fig:TreeChannels}. 
\begin{figure*}[t]
	\includegraphics[scale=1.15]{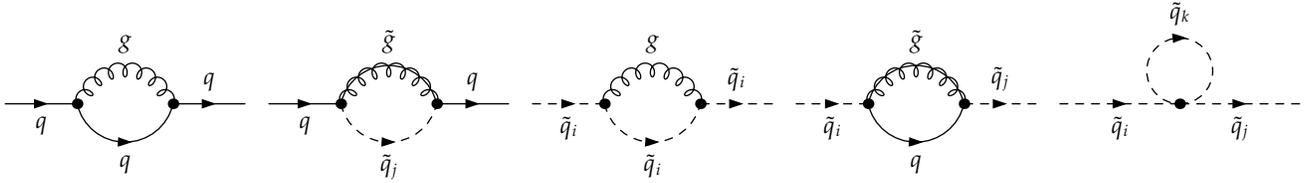}
	\caption{Self-energy corrections for the quarks and squarks at one-loop level in QCD contributing to neutralino-squark co-annihilation.}
	\label{Fig:SelfEnergies}
\end{figure*}
\begin{figure*}[t]
	\includegraphics[scale=0.92]{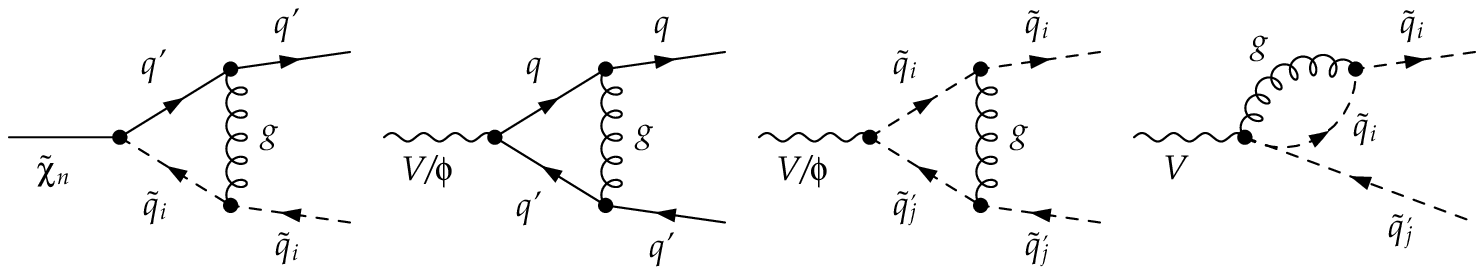}
	\includegraphics[scale=0.92]{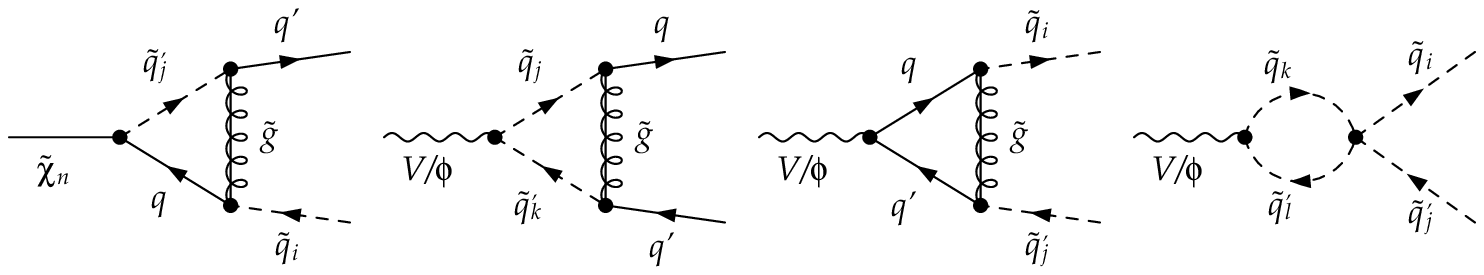}
	\caption{Vertex corrections at one-loop level contributing to neutralino-squark co-annihilation 
	into quarks and Higgs ($\phi$) or electroweak gauge ($V$) bosons. The diagram involving the $V-g-\tilde{q}-\tilde{q}$ vertex is present only for the case of a gauge boson in the final state.}
	\label{Fig:VertexCorrections}
\end{figure*}

In comparison to the first parameter point, scenario II has a smaller
$\mu$-parameter, but a larger value of $\tan\beta$. Moreover, the gauginos
and third-generation squarks are slightly heavier, and the trilinear coupling
is slightly lower than for scenario I. As a consequence, the relative
importance of the co-annihilation channels is altered, as can be seen in
Tab.\ \ref{Tab:Channels}. In particular, the co-annihilation into the lightest
Higgs boson contributes less, allowing the final state containing a $Z$-boson
to become more important. In contrast to the co-annihilation into a Higgs
boson, the dominant diagram in this case is the exchange of a top quark in the
$s$-channel, as can be seen in Fig.~\ref{Fig:TreeChannels} (lower left plot). 
For this scenario, we also show the individual contributions of the three
diagrams for co-annihilation into a bottom quark and a $W$-boson (lower right
plot). As in the previous case, the $s$-channel is the dominant mode. Its
absolute cross section value is even larger than for $tZ^0$ due to the larger
phase space. However, large destructive interferences of this diagram with
the sub-leading $t$- and $u$-channels decrease its cross section, so that the
total value is almost an order of magnitude smaller than for the Z-boson.

Finally, scenario III is quite similar to scenario I with the exception of
a very light pseudoscalar Higgs boson of $m_{A^0} = 386$ GeV. This leads to
a similarly light heavy $CP$-even Higgs boson $H^0$ (see Tab.\
\ref{Tab:Scenarios}). As a consequence, the co-annihilation into heavy CP-even
Higgs bosons in association with a top quark is now open and becomes the
dominant contribution to neutralino-stop co-annihilation (see Tab.\
\ref{Tab:Channels}). The final state containing a light Higgs boson remains
important, while co-annihilations into $Z$- and $W$-bosons are marginal for
this parameter point. As it was the case for the lightest Higgs boson, the
co-annihilation into $t H^0$ is dominated by the exchange of a scalar top in
the $t$-channel (upper right plot of Fig.~\ref{Fig:TreeChannels}),
which is again due to the enhanced trilinear coupling. The dominance is even
more important here, which is explained by the modified mixing in the Higgs
sector due to the smaller mass difference between $h^0$ and $H^0$.
%

\section{One-loop cross sections \label{Sec:Calculation}}
\begin{figure*}[t]
	\includegraphics[scale=1.15]{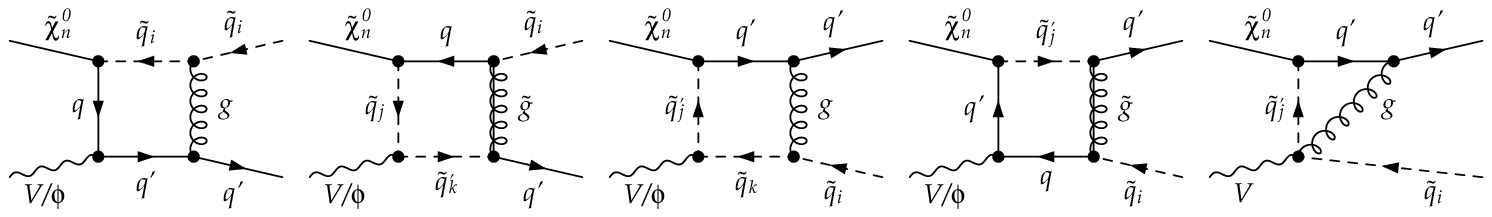}
	\caption{Four-point diagrams at one-loop level contributing to neutralino-squark co-annihilation into quarks and Higgs ($\phi$) or electroweak gauge ($V$) bosons. The last diagram involving the four-vertex is absent for a scalar in the final state.}
	\label{Fig:BoxDiagrams}
\end{figure*}

In this section, we turn to a detailed discussion of our analytical calculations
of the full QCD and SUSY-QCD corrections to neutralino-stop co-annihilation into
electroweak gauge and Higgs bosons. We first describe the computation of the
virtual loop corrections and the renormalization scheme employed in the quark
and squark sector, then the analytical evaluation of real gluon emission
diagrams and the corresponding cancellation of infrared singularities with those
encountered in the virtual contributions. Finally, we address the subtle point
of intermediate on-shell particles and how we subtract their resonant
contributions.

\subsection{Virtual corrections and renormalization \label{Sec:RenScheme}}
The co-annihilation processes considered in this paper (see Fig.~\ref{Fig:TreeDiagrams}) 
include strongly interacting particles both in the initial and final states. As a consequence, 
the leading higher-order corrections to these processes come from loop diagrams containing a 
gluon, a gluino, a four-squark vertex, and from real radiation processes when a gluon is 
emitted from one of the squarks or quarks. The virtual corrections for the co-annihilation 
processes contain propagator corrections, vertex corrections, and box diagrams. The corresponding 
diagrams are shown in Figs.~\ref{Fig:SelfEnergies}, \ref{Fig:VertexCorrections}, and \ref{Fig:BoxDiagrams}, 
respectively. The divergences in these diagrams are regularized by performing the calculations 
in $D=4-2\varepsilon$ dimensions. In order to preserve supersymmetry in the process, we use the 
dimensional reduction regularization scheme ($\overline{\tt DR}$). All tensor loop integrals are 
reduced using the standard Passarino-Veltman reduction \cite{Passarino:1978jh}. The resulting scalar 
integrals are evaluated using the known results in, e.g., Refs.~\cite{Dittmaier:2003bc, Denner:2010tr}. 
The renormalization and factorization scales are set to the center-of-mass energy $\sqrt{s}$.
The amplitudes corresponding to the Feynman diagrams in Figs.~\ref{Fig:SelfEnergies}-\ref{Fig:BoxDiagrams} 
have been calculated analytically and cross-checked using the publicly available tools {\FeynArts} 
\cite{FeynArts}, {\FeynCalc} \cite{FeynCalc}, and {\FORM} \cite{FORM}. 

In order to cancel all arising ultraviolet (UV) singularities and render the
cross section UV-finite, we introduce counterterms to the relevant MSSM
parameters and fields. When considering QCD and SUSY-QCD corrections to all 
processes that are needed to determine the neutralino relic density, a consistent 
treatment of all parameters in the quark and squark sector of the MSSM is essential. 
We introduce a hybrid on-shell/$\overline{\tt DR}$ renormalization scheme which 
is set up in such a way that it minimizes potential problems connected to sensitive 
parameters (e.g.\ the bottom trilinear coupling $A_b$) and is valid in a large region 
of MSSM parameter space. We will introduce all parameters and their treatment in 
detail below. We have explicitly verified that after renormalization all
UV divergences cancel. The resulting expressions together with the
renormalization scheme have been implemented in a numerical fortran code
\cite{DMNLO}, which can serve as an extension to public dark matter tools
such as {\DS} \cite{DarkSusy} and {\MO} \cite{micrOMEGAs2007}. Similar renormalization 
schemes for the quark and squark sectors of the MSSM were already introduced and studied 
in Refs.~\cite{Baro:2009gn, Heinemeyer:2010mm}. Compared to those analyses, our approach 
differs significantly in the treatment of the squark mixing angles $\theta_b$ and $\theta_t$, 
but shares some important features with the RS2 scheme introduced in Ref.~\cite{Heinemeyer:2010mm}.

\subsubsection{Quark sector} 

The process of neutralino-stop co-annihilation considered here involves
only quarks and squarks of the third generation. We will therefore discuss
only the case of massive quarks. The parameters to be renormalized are
the quark fields and masses. We perform the wave-function renormalization
by introducing counterterms $\delta Z_{L,R}$ for each chirality of the
third-generation quarks
\begin{equation}
\left(\begin{array}{c}
 q_L \\ q_R
      \end{array}\right) \rightarrow \left(\begin{array}{cc}
1+\frac{1}{2}\delta Z_L & 0 \\ 0 & 1+\frac{1}{2}\delta Z_R
     \end{array}\right)\left(
\begin{array}{c}
 q_L \\ q_R
     \end{array}\right)\,.
\end{equation}
The wave-function renormalization constants are fixed by requiring
the external quark propagators to have unit residue even at
one-loop order. This leads to the following expression for the massive quarks ($q=t,b$)
\begin{align}\nonumber
	\delta Z_L = \Re\,\Big\{ &-\Pi_{L} (m_q^2)- m_q^2\Big[ \dot\Pi_{L}
	(m_q^2)+ \dot\Pi_{R} (m_q^2)\Big]\\ \nonumber
	 &+ \frac{1}{2 m_q}\Big[\Pi_{SL}(m_q^2)-\Pi_{SR} (m_q^2)\Big] \\ \label{offdiag}
	 &- m_q\Big[\dot\Pi_{SL} (m_q^2)+\dot\Pi_{SR}
	(m_q^2)\Big]\Big\}\,,
	\\ 
	\delta Z_R = \delta Z_L &(L \leftrightarrow R)\,,
\end{align}
where $\Pi_{L,R}(k^2)$ and $\Pi_{SL,SR}(k^2)$ stand for the vector and the scalar parts of the 
two-point Green's function as defined in Ref.~\cite{Kovarik2005} and 
$\dot\Pi (m^2)=\left[\frac{\partial}{\partial k^2}\Pi(k^2)\right]_{k^2=m^2 }$. 

After the wave-function renormalization has been performed, we still
have to renormalize the masses of the quarks. Although both the top and
bottom quark are heavy, their properties are very different, and so
is our treatment of their masses. On the one hand, the top quark does not 
form bound states and its physical mass is directly measurable. Therefore 
in our calculation, we use the physical (on-shell) top quark mass 
$m_t=173.1\ {\rm GeV}$. This implies using the on-shell mass counterterm 
for the top quark defined as
\begin{align}
	\delta m_{t} = \frac{1}{2}\, \Re \Big\{ & m_{t}
	\Big[ \Pi_L (m_{t}^2) + \Pi_R (m_{t}^2) \Big]\\ \nonumber & +
	\Pi_{SL}(m_{t}^2) + \Pi_{SR}(m_{t}^2) \Big\} \,.
\end{align}

On the other hand, the bottom quark forms hadrons and its mass cannot be 
directly measured. Conventionally a mass parameter $m_b(m_b)$ is extracted 
in the $\overline{\tt MS}$ renormalization scheme from the Standard Model 
analysis of $\Upsilon$ sum rules \cite{MBmass}. In order to obtain the appropriate 
bottom quark mass in the $\overline{\tt DR}$ renormalization scheme within the MSSM, 
we first use the Standard Model next-to-next-to-leading order (NNLO) renormalization 
group evolution to obtain the mass of the bottom quark at a scale $Q$ \cite{Baer2002}. 
We then convert the $\overline{\tt MS}$ mass $m_b^{\overline{\tt MS},\,{\tt SM}}(Q)$ 
to a mass in the $\overline{\tt DR}$ renormalization scheme 
$m_b^{\overline{\tt DR},\,{\tt SM}}(Q)$ while still in the Standard Model \cite{Baer2002}. 
Finally we apply the threshold corrections including also contributions from SUSY particles 
in the loop (denoted by $\Delta m_b$)
\begin{equation}
        m_{b}^{\overline{\tt DR},\,{\tt MSSM}}(Q) ~=~ m_{b}^{\overline{\tt DR},\,{\tt SM}}(Q) - \Delta m_b\, .
\end{equation}
The corresponding counterterm contains the pole in $\varepsilon
= (4-D)/2$ and can be written as
\begin{equation}\label{Eq:quarkDR}
	\frac{\delta m_b^{\overline{\tt DR}}}{m_b} = (-2)\frac{\alpha_s C_F}{4\pi} \frac{c_\varepsilon}{\varepsilon}\,,
\end{equation}
where we factored out the constant $c_\varepsilon = \Gamma(1+\varepsilon)(4\pi)^{\varepsilon}$. 
One prominent place where the quark masses enter the calculation is through the Yukawa couplings 
of the Higgs bosons to the quarks. Especially the Yukawa couplings of the bottom quark were extensively 
studied in the decays of Higgs bosons in the Standard Model. Important QCD and top-quark induced 
corrections to the coupling of Higgs bosons to bottom quarks were calculated up to 
$\mathcal{O}(\alpha_s^4)$ \cite{QCDhiggs} and can be used to define an effective Yukawa coupling which
includes these corrections as
\begin{equation}
\big[\big(h_b^{\overline{\tt MS},{\tt QCD},\Phi}\big)(Q)\big]^2 = 
\big[\big(h_b^{\overline{\tt MS},\Phi}\big)(Q)\big]^2 \Big[1 + \Delta_{\tt QCD} + \Delta_{t}^\Phi\Big]\,,
\end{equation}
for each Higgs boson $\Phi=h^0,H^0,A^0$. The QCD corrections $\Delta_{\tt QCD}$ are explicitly given by
\begin{align}\nonumber
	\Delta_{\tt QCD} &\ \;= \frac{\alpha_s(Q)}{\pi} C_F\frac{17}{4} + \frac{\alpha^2_s(Q)}{\pi^2}\Big[ 35.94 - 1.359\, n_f \Big]\\
	& + \frac{\alpha^3_s(Q)}{\pi^3}\Big[164.14 - 25.77\, n_f + 0.259\,n^2_f\Big]\\ \nonumber & + 
	\frac{\alpha^4_s(Q)}{\pi^4}\Big[39.34 - 220.9\,n_f +9.685\,n^2_f - 0.0205\, n^3_f\Big]\,,
\end{align}
and the top-quark induced corrections $\Delta_{t}^\Phi$ for each Higgs boson $\Phi$ read
\begin{align}
	\Delta_{t}^h = c_h(Q)&\Biggr[ 1.57 - \frac23 \log \frac{Q^2}{m_t^2}+ 
	\frac19 \log^2\frac {m^2_b(Q)}{Q^2} \Biggr],\\
	\Delta_{t}^H = c_H(Q)&\Biggr[ 1.57 - \frac23 \log \frac{Q^2}{m_t^2}+ 
	\frac19 \log^2\frac {m^2_b(Q)}{Q^2} \Biggr],\\
	\Delta_{t}^A = c_A(Q)&\Biggr[ \frac{23}{6} - \log \frac{Q^2}{m_t^2}+ 
	\frac16 \log^2\frac {m^2_b(Q)}{Q^2} \Biggr]\,,
\end{align}
with
\begin{multline}
	\big\{c_h(Q),c_H(Q),c_A(Q)\big\} = \\ \frac{\alpha_s^2(Q)}{\pi^2}
	\Big\{\frac{1}{\tan\alpha\tan\beta}, 
	      \frac{\tan\alpha}{\tan\beta},  
	      \frac{1}{\tan^2\beta}\Big\}\,.
\end{multline}
We take into account these corrections excluding the one-loop part as it is 
provided consistently through our own calculation.

In the MSSM, the Yukawa coupling to bottom quarks can receive large corrections for 
large $\tan\beta$ or large $A_b$, even beyond the next-to-leading order, which can affect 
our analysis. Therefore, in addition, we include these corrections that can be resummed 
to all orders in perturbation theory \cite{Carena2000,Spira2003}. Denoting the 
resummable part by $\Delta_b$ we redefine the bottom quark Yukawa couplings as
\begin{eqnarray}
	h_b^{{\tt MSSM},h}(Q) &=& \frac{h_b^{\overline{\tt MS},{\tt QCD},h}(Q)}{1+\Delta_b}
	\Biggr[1-\frac{\Delta_b}{\tan\alpha\tan\beta}\Biggr]\,, \\
	h_b^{{\tt MSSM},H}(Q) &=& \frac{h_b^{\overline{\tt MS},{\tt QCD},H}(Q)}{1+\Delta_b}
	\Biggr[1+\Delta_b \frac{\tan\alpha}{\tan\beta}\Biggr]\,, \\
	h_b^{{\tt MSSM},A}(Q) &=& \frac{h_b^{\overline{\tt MS},{\tt QCD},A}(Q)}{1+\Delta_b}
	\Biggr[1-\frac{\Delta_b}{\tan^2\beta}\Biggr]\,.
\end{eqnarray}
In the same way as for the QCD corrections, we exclude the one-loop
part of these SUSY-QCD corrections and include only the resummed
remainder, since the one-loop part is already present in our calculation.

\subsubsection{Squark sector}

As in the above discussion for quarks, we will address here only the
squarks of the third generation, i.e.\ stops and sbottoms. We work in
the mass eigenstate basis and introduce the wave-function
renormalization counterterms $\delta Z_{ij}$ through
\begin{equation}
	\tilde{q}_i \rightarrow \Big(\delta_{ij}+\frac{1}{2}\delta Z_{ij}\Big)\tilde{q}_j\,,
\end{equation}
where in contrast to the case of quarks the $\delta Z_{ij}$ include
also off-diagonal terms. The wave-function renormalization counterterms
are again fixed by requiring that the squark propagators have unit
residue also at one-loop level. In addition we require that mixing for
on-shell squarks is absent. These conditions lead to the counter\-terms
\begin{align}
	\delta Z_{ii} &= - \Re\,\Big[\dot\Pi_{ii}^{\tilde{q}} (m_{\tilde{q}_i}^2)\Big]\,,\\ 
	\delta Z_{ij} &= \frac{2\,\Re\,\big[\Pi_{ij}^{\tilde{q}}(m_{\tilde{q}_j}^2)\big]}{m_{\tilde{q}_{i}}^2-m_{\tilde{q}_{j}}^2}
	,\qquad {\rm for}\quad i\neq j\,,
\end{align}
where $\Pi_{ij}^{\tilde{q}}(k^2)$ are again the two-point Green's functions,
this time for squarks.

The renormalization of the squark masses is complicated due to the mixing of 
squarks of the third generation. Therefore, it has to be discussed in conjunction 
with the renormalization of all other parameters in the squark sector appearing 
in the mass matrix. At tree-level, the masses $m^2_{\tilde{q}_i}$ for stops 
and sbottoms are obtained by diagonalization of the mass matrix
\begin{widetext}
\begin{equation}
	U^{\tilde q} 
	\left( \begin{array}{cc} M_{\tilde Q}^2
	       + (I^{3L}_q \!-\! e_q\,s_W^2)\cos2\beta\,
	       m_{Z}^{\,2}
	       + m_{q}^2\, & m_q\big( A_q - \mu \,(\tan\beta)^{-2 I^{3L}_q}\big) \\  
	m_q \big( A_q - \mu \,(\tan\beta)^{-2 I^{3L}_q}\big) & M_{\{\tilde U,\,\tilde D\}}^2
		       + e_{q}\,s_W^2 \cos2\beta\,m_{Z}^{\,2}
		       + m_q^2 \end{array} \right) 
	(U^{\tilde q})^\dag
	~=~
	\left( \begin{array}{cc} m^2_{\tilde{q}_1} & 0 \\ 0 & m^2_{\tilde{q}_2} \end{array} \right)	,\label{Eq:massMatrix}
\end{equation}
\end{widetext}
where $e_q$ is the fractional charge of the squark in units of $e$, $s_W$
is the sine of weak mixing angle, $I^{3L}_q$ is the weak isospin of the squark,
and $U^{\tilde q}$ are the squark mixing matrices. As it is well known, we have to
consider both the stop and the sbottom sector at the same time, since due to
$SU(2)$ symmetry the mass matrices share a common soft breaking parameter
$M_{\tilde Q}^2$ connecting the two sectors. In fact, out of the total set of
eleven parameters $M_{\tilde Q}^2,M_{\tilde U}^2,M_{\tilde D}^2,A_{t},A_{b},
\theta_{\tilde t},\theta_{\tilde b},m^2_{\tilde{t}_1},m^2_{\tilde{t}_2},m^2_{\tilde{b}_1}$,
and $m^2_{\tilde{b}_2}$, only five are completely independent and can be
considered as input parameters. Their counterterms can then be freely chosen.
The remaining parameters are derived by requiring that Eq.\
(\ref{Eq:massMatrix}) is valid even at one-loop order. 

Here, we adopt a hybrid on-shell/$\overline{\tt DR}$ renormalization scheme
choosing as input the parameters $A_{t},A_{b},m^2_{\tilde{t}_1},m^2_{\tilde{b}_1}$,
and $m^2_{\tilde{b}_2}$, where the trilinear couplings $A_{t},A_{b}$ are defined
in the $\overline{\tt DR}$ renormalization scheme and all input masses are
defined on-shell. This choice is motivated by the fact that we want to
obtain a renormalization scheme which is applicable for all annihilation and
co-annihilation processes, where squarks play an important role. For example,
as the co-annihilation processes are extremely sensitive to the mass of the
lightest stop and as this mass also plays an important role in the $t$-channel
exchange of neutralino annihilations \cite{DMNLO_NUHM}, we choose to include
its mass in the input parameters. It is then crucial to take its
physical/on-shell definition. Moreover, due to the appearance of the trilinear
parameters $A_{t},A_{b}$ in the important Higgs-squark-squark
coupling in the co-annihilation processes, it is a natural choice to include
them in our input set as well. Given the possible problems with the one-loop
definition of the $A_{b}$ parameter widely discussed in the literature
\cite{Heinemeyer:2010mm, Eberl:1999he, Weber:2007id}, we choose to define
both trilinear parameters in the $\overline{\tt DR}$ scheme. A different
approach would be to define these parameters in the on-shell scheme, e.g.\
through the decay process of a squark into a squark and a Higgs boson
\cite{Baro:2009gn}. This, however, would require a dedicated treatment of the
infrared divergences arising in such a calculation.

Having explained above our choice of renormalization scheme, we must now
specify the counterterms for the input parameters depending on their definition.
The counterterms for the on-shell masses $m^2_{\tilde{t}_1},m^2_{\tilde{b}_1}$, and
$m^2_{\tilde{b}_2}$ are defined in the usual way as
\begin{equation}
	\delta m_{\tilde{q}_i}^2 = \Re\, \Big[\Pi_{ii}^{\tilde{q}}(m_{\tilde{q}_i}^2)\Big]\,. 
\end{equation}
The $\overline{\tt DR}$ counterterms of the trilinear parameters contain
only the UV poles and can be given in terms of other $\overline{\tt DR}$
counterterms as
\begin{align}\nonumber
	\delta A_{\tilde{q}}^{\overline{\tt DR}} = &\frac{1}{m_q}\Biggr[ 
 U^{\tilde{q}}_{11}U^{\tilde{q}}_{12}(\delta m^2_{\tilde{q}_1})^{\overline{\tt DR}}
+ U^{\tilde{q}}_{21}U^{\tilde{q}}_{22}(\delta m^2_{\tilde{q}_2})^{\overline{\tt DR}} \\ \nonumber
&\ \ + \big(U^{\tilde{q}}_{21}U^{\tilde{q}}_{12} +
U^{\tilde{q}}_{11}U^{\tilde{q}}_{22}\big)\big(m^2_{\tilde{q}_1}-m^2_{\tilde{q}_2}\big)
\delta\theta_{\tilde{q}}^{\overline{\tt DR}} \\
&\ \ - \frac{\delta m_q^{\overline{\tt DR}}}{m_q}\big(U^{\tilde{q}}_{11}U^{\tilde{q}}_{12} m^2_{\tilde{q}_1} + U^{\tilde{q}}_{21}U^{\tilde{q}}_{22} m^2_{\tilde{q}_2}\big)\Biggr]\,.
\end{align}
The remaining $\overline{\tt DR}$ counter\-terms for squark masses and their mixing angle 
are given as (for $j \neq i$; for the quark mass counterterm see Eq.~(\ref{Eq:quarkDR}))
\begin{widetext}
\begin{eqnarray}
	(\delta m^2_{\tilde{q}_i})^{\overline{\tt DR}} &=& \frac{\alpha_s C_F}{4\pi} \frac{c_\varepsilon}{\varepsilon}
	\Biggr[ \big( (U^{\tilde{q}}_{i1})^2-(U^{\tilde{q}}_{i2})^2\big)^2 m^2_{\tilde{q}_i} -m^2_{\tilde{q}_i}
	 \ + \big(U^{\tilde{q}}_{21} U^{\tilde{q}}_{11} - U^{\tilde{q}}_{22} U^{\tilde{q}}_{12} \big)^2 m^2_{\tilde{q}_j} 
     \ + 8 m_q m_{\tilde{g}}\, U^{\tilde{q}}_{i1} U^{\tilde{q}}_{i2}  - 4 m^2_{\tilde{g}} - 4 m_q^2 \Biggr]\,, \nonumber \\
	\delta\theta_{\tilde{q}}^{\overline{\tt DR}} &=& \frac{\alpha_s C_F}{4\pi}\frac{c_\varepsilon}{\varepsilon} \frac{1}{(m^2_{\tilde{q}_1}-m^2_{\tilde{q}_2})}\, \Biggr[\big(U^{\tilde{q}}_{21} U^{\tilde{q}}_{11} - U^{\tilde{q}}_{22} U^{\tilde{q}}_{12} \big)\Big( \big( (U^{\tilde{q}}_{11})^2-(U^{\tilde{q}}_{12})^2\big)^2 m^2_{\tilde{q}_1} + \big( (U^{\tilde{q}}_{21})^2-(U^{\tilde{q}}_{22})^2\big)^2 m^2_{\tilde{q}_2}\Big) \nonumber \\ 
	& & + 4m_{\tilde{g}}m_q \big(U^{\tilde{q}}_{11}U^{\tilde{q}}_{22} + U^{\tilde{q}}_{12}U^{\tilde{q}}_{21}\big) \Biggr]\,.
\end{eqnarray}
\end{widetext}

The values of the dependent parameters $M_{\tilde Q}^2,M_{\tilde U}^2,M_{\tilde D}^2,m^2_{\tilde{t}_2}, 
\theta_{\tilde t}$, and $\theta_{\tilde b}$ are determined using
Eq.~(\ref{Eq:massMatrix}). For example, by taking a trace and a determinant of both
sides of Eq.~(\ref{Eq:massMatrix}) for stops and sbottoms, we can relate the four
parameters $M_{\tilde Q}^2,M_{\tilde U}^2,M_{\tilde D}^2$, and $m^2_{\tilde{t}_2}$ to the on-shell
sfermion masses and the other parameters of the mass matrix such as $\mu$ or
$\tan\beta$, which do not receive any QCD corrections and hence do not require
renormalization. Having determined all mass parameters, we diagonalize the stop and sbottom mass
matrices leading to the values of both mixing matrices. The eigenvalues are then the
chosen on-shell masses and by construction the dependent mass $m^2_{\tilde{t}_2}$.

The counterterms of the dependent parameters are derived also from the defining Eq.~(\ref{Eq:massMatrix}). 
We do not give counterterms for $M_{\tilde Q}^2,M_{\tilde U}^2,M_{\tilde D}^2$ as they never appear in any vertex. 
Unlike in other analyses where the mixing angles are the input parameters and their counterterms are, e.g., 
given as a combination of wave-function renormalization constants \cite{Kovarik2005}, here both mixing angles 
$\theta_{\tilde t}$ and $\theta_{\tilde b}$ are dependent and have the counterterms
\begin{widetext}
\begin{equation}
	\delta\theta_{\tilde q} = \frac{1}{\big(U^{\tilde{q}}_{21}U^{\tilde{q}}_{12} + 
	U^{\tilde{q}}_{11}U^{\tilde{q}}_{22}\big)\big(m^2_{\tilde{q}_1}-m^2_{\tilde{q}_2}\big)}
    \Big(\delta m_q \big( A_q - \mu \,(\tan\beta)^{-2 I^{3L}_q}\big) + m_q\, \delta A_q - 
    U^{\tilde{q}}_{11}U^{\tilde{q}}_{12}\big(\delta m^2_{\tilde{q}_1}-\delta m^2_{\tilde{q}_2}\big)\Big)\,.
\end{equation}
In the case of the stop mixing matrix this counterterm includes the last remaining undetermined 
counterterm of the mass of the heavy stop quark
\begin{eqnarray}\nonumber
	\delta m^2_{\tilde{t}_2} = \frac{1}{U^{\tilde{t}}_{21}U^{\tilde{t}}_{12}}
	&&\Bigg[\big(U^{\tilde{t}}_{21}U^{\tilde{t}}_{12} + 
	U^{\tilde{t}}_{11}U^{\tilde{t}}_{22}\big) 
	\Big((U^{\tilde{b}}_{11})^2\delta m^2_{\tilde{b}_1} + (U^{\tilde{b}}_{21})^2\delta m^2_{\tilde{b}_2} + 
	2 U^{\tilde{b}}_{11}U^{\tilde{b}}_{21}\big(m^2_{\tilde{b}_1}-m^2_{\tilde{b}_2}\big)\delta\theta_{\tilde{b}}
	-2 m_b\delta m_b \\
	&& - (U^{\tilde{t}}_{11})^2\delta m^2_{\tilde{t}_1} + 2 m_t\delta m_t\Big) - 
	2 U^{\tilde{t}}_{11} U^{\tilde{t}}_{21}\Big(\delta m_t \big( A_t - \mu/\tan\beta\big) + 
	m_t\, \delta A_t - U^{\tilde{t}}_{11}U^{\tilde{t}}_{12}\delta m^2_{\tilde{t}_1}\Big)\Bigg]\,.
\end{eqnarray}
\end{widetext}

This concludes our discussion of our renormalization scheme. We have discussed in detail the 
definition and renormalization of every relevant parameter in the quark and squark sector. By a clever 
choice of parameters we obtain a renormalization scheme which works in large parts of the relevant parameter 
space of the MSSM for all annihilation and co-annihilation processes where quarks and squarks play an crucial role.  

\subsection{Real corrections and infrared treatment}

\begin{figure*}
	\includegraphics[scale=1.15]{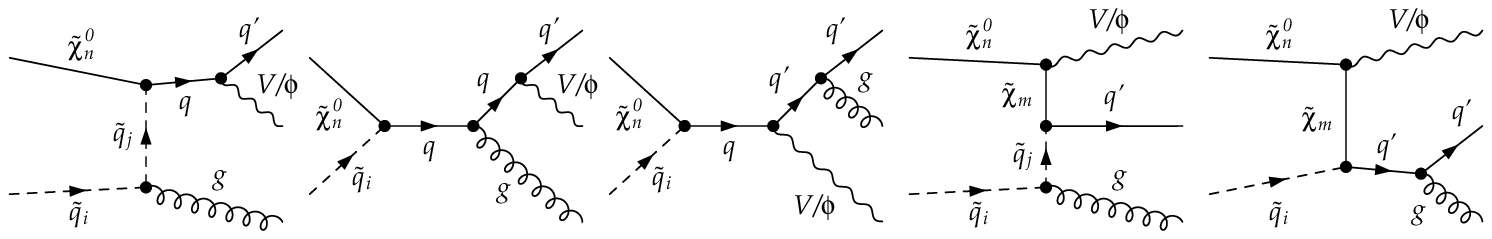}
	\includegraphics[scale=0.92]{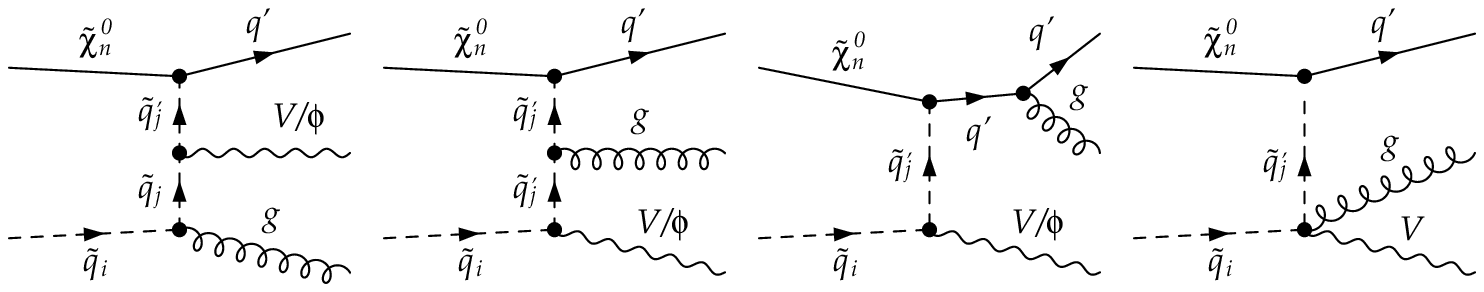}
	\caption{Real gluon emission diagrams at one-loop level contributing to neutralino-squark 
	co-annihilation into quarks and Higgs ($\phi$) or electroweak gauge ($V$) bosons. The last diagram involving the four-vertex is absent for a scalar in the final state.}
	\label{Fig:RealEmission}
\end{figure*}
\begin{figure*}[t]
	\includegraphics[scale=0.5]{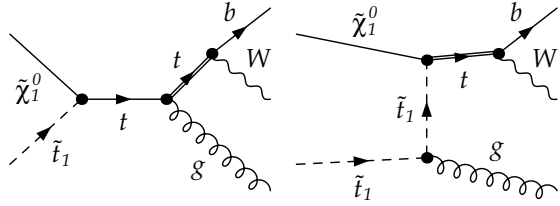}
	\caption{Real gluon emission diagrams with a $Wb$ final state where an internal top quark can become on-shell, 
	as indicated by a double line.}
	\label{Fig:OnShell}
\end{figure*}

Including only the virtual corrections with the renormalization constants does not lead to a finite result as some 
diagrams where a gluon is exchanged lead to a different type of divergence - the infrared (IR) divergence. 
These divergences cancel against similar
divergences that come from the real radiation corrections where a gluon is emitted from one of the quarks or squarks, 
see diagrams in Fig.~\ref{Fig:RealEmission}. The cancellation of these divergences is not as straightforward as in 
the case of ultraviolet divergences discussed above. It is because the IR divergence in the virtual diagrams can be 
explicitly isolated again by working in a general dimension $D$, whereas the divergence in the real corrections comes 
from the phase-space integration over the gluon phase-space.

Several approaches exist in order to cancel these divergences, most notably the so-called phase-space slicing method
\cite{HarrisOwens,GieleGlover,Denner:1991kt} or the dipole subtraction method \cite{Catani-Seymour}.\footnote{The 
implementation of a dedicated dipole subtraction method \`a la Catani-Seymour \cite{Catani-Seymour} is work in 
progress and subject to a later publication.} Here we use the phase-space slicing
method which uses a lower cut on the gluon energy $\Delta E$ in the phase-space integration to render the real 
corrections finite. The missing
divergent piece of the phase-space integral can be performed analytically in the limit of small energy of the gluon - the so-called soft-gluon approximation. Divergences obtained in the soft-gluon approximation then cancel analytically with those coming from the virtual 
corrections.
In the soft-gluon approximation the phase-space integration factorizes as
\begin{equation}
\left( \frac{\dd\sigma}{\dd\Omega} \right)_{\textnormal{soft}} = F \times \left( \frac{\dd\sigma}{\dd\Omega} \right)_{\textnormal{tree-level}},
\end{equation}
where $F$ contains the integral over the phase-space of the gluon and therefore also the divergence. Explicitly, $F$ contains integrals of the form
\begin{equation}
I_{ab} =  \mu^{4-D}\int_{|\vec{k}| \le \Delta E} \frac{\dd^{D-1} k}{(2\pi)^{D-4}}\frac{1}{k^0} \frac{(a.b)}{(k.a)(k.b)},
\label{Eq:SoftIntegral}
\end{equation}
where $k$ is the $4-$momentum of the gluon and $a$ and $b$ are $4-$momenta of two external particles which can emit a gluon. 
These integrals are given in Ref.\ \cite{Denner:1991kt, tHooft:1978xw}. In our case we use dimensional regularization 
to obtain an explicit form of the divergence.

The phase-space slicing method introduces a cut $\Delta E$ to separate the divergent part of the phase-space. It appears in the
original real corrections as a lower limit on the integration over the energy of the gluon and also explicitly in the cross section 
calculated in the soft-gluon approximation. In principle the dependence on this cut should completely vanish, but in 
practice the cancellation is limited by the stability of numerical integration of the real corrections. For practical 
purposes one has to choose a value for the cut such that it is small enough for the soft-gluon approximation to be valid in 
the region of phase-space given by $|\vec{k}| \le \Delta E$, but at the same time large enough for the numerical integration of the real 
correction to be still possible. We have verified that in our calculation all cross sections are insensitive to the choice of this cut.

\subsection{On-shell propagators}
While including next-to-leading order corrections to the studied neutralino co-annihilation processes, we have to take care of a few subtleties. Some processes, although well defined and separate at tree-level, cannot be 
unambiguously defined and separated when NLO corrections are considered. One such example is the process 
$\tilde{\chi}_1^0 \tilde{t}_1 \to b W$. Here, additional gluon radiation can be taken to be a 
real correction to the $Wb$ process. However, it can equally well be considered to be neutralino-stop co-annihilation with a gluon and a top quark in the final state where the top decays into a $W$-boson and a bottom quark.
Despite the fact that these processes cannot be separated at NLO and one should strictly speaking include also their interference, 
for practical purposes it is desirable to find a way how to separate them.
\begin{figure*}[t]
	\includegraphics[scale=0.43]{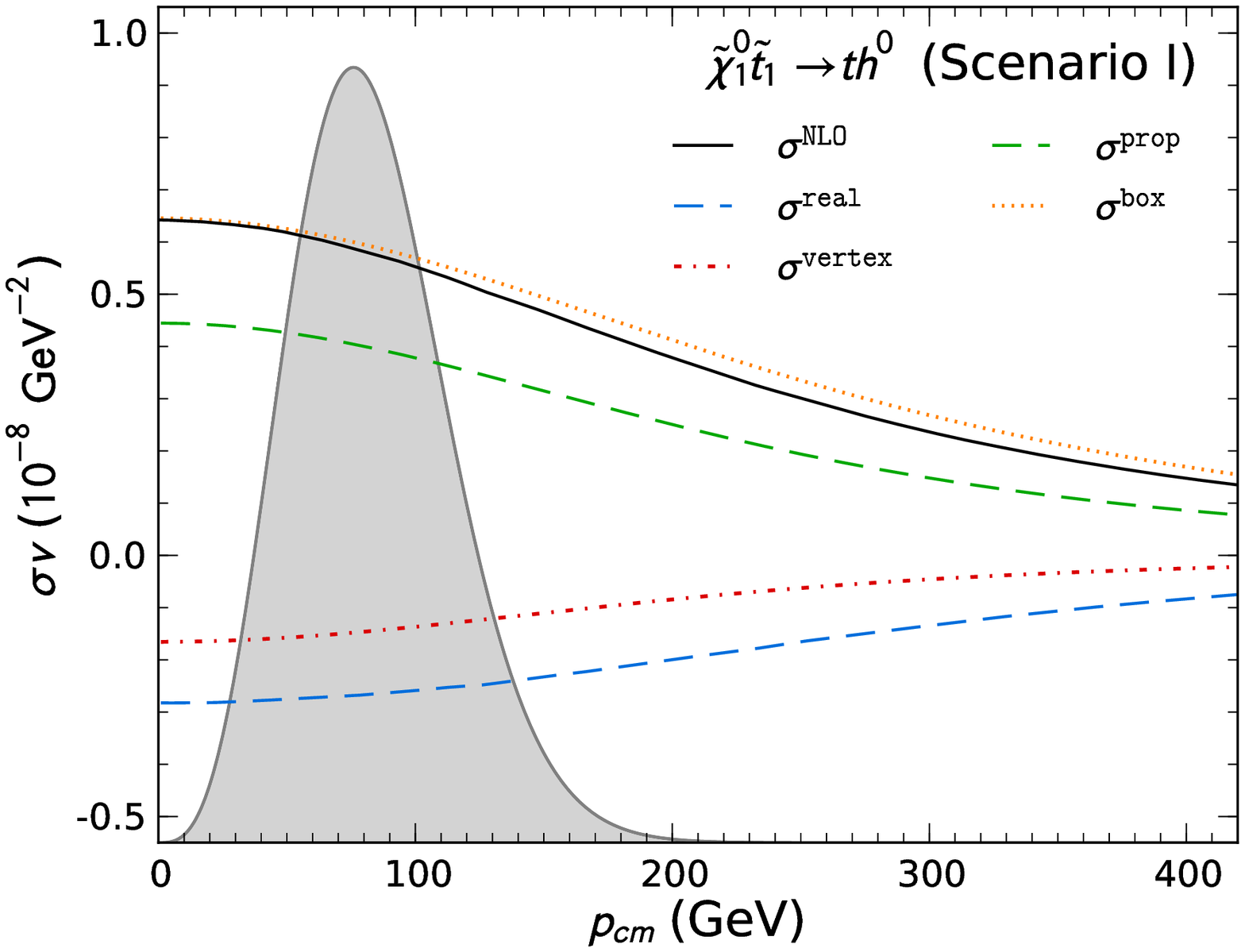}
	\includegraphics[scale=0.43]{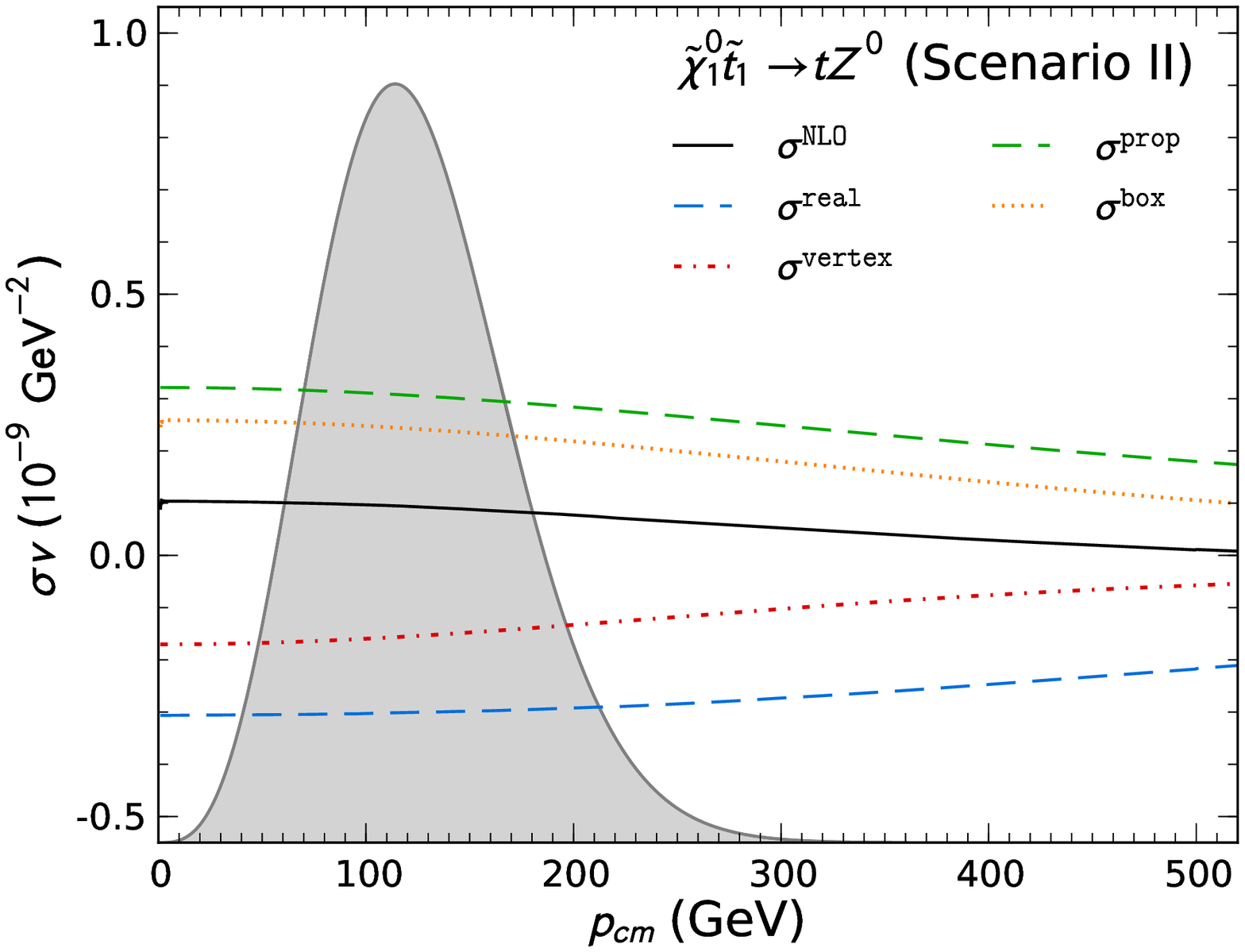}\\
	\caption{Contribution of the different corrections to the total next-to-leading order correction for the 
	case of co-annihilation into $t h^0$ for scenario I and into $t Z^0$ for scenario II. The real contribution $\sigma^{\rm real}$ is defined as the sum of the hard radiation and
	the soft gluon part with a cut on the gluon energy of $\Delta E = 10^{-3}
	\sqrt{s}$. The gray area indicates the thermal distribution (in arbitrary units).}
	\label{Fig:CorrContrib}
\end{figure*}

Due to the above mentioned complication, one has to treat the process $\tilde{\chi}_1^0 \tilde{t}_1 \to b W g$ with care as it contains 
a top quark propagator which can become on-shell. At tree-level the large masses of the neutralino and the scalar top 
quark prevent the internal top quark to be on-shell. In contrast, when an additional gluon is radiated either from the initial stop or the internal 
top-quark propagator, the gluon can carry away enough energy for the top propagator to become on-shell. The relevant diagrams where this 
can occur are shown in Fig.~\ref{Fig:OnShell}. 
We regularize the appearing divergence from the on-shell propagator by introducing a width $\Gamma_t$ for the top quark in the 
problematic propagators, leading to a finite result for the integrated matrix elements for the real gluon emission. 
The matrix element when integrated over the whole phase-space is very large as it includes also the leading order 
co-annihilation process $\tilde{\chi}_1^0 \tilde{t}_1 \to t g$ with the top quark decaying into $W^+ b$. 
This process is, however, already accounted for in the calculation of the neutralino relic density. To avoid double-counting, we need to separate the two processes.

In order to treat the double-counting in the real correction contribution, we use a local on-shell subtraction scheme
 \cite{Beenakker1997,Tait:1999cf,GoncalvesNetto:2012yt}, in which a locally gauge invariant term is subtracted from 
the original cross section that has been regularized as discussed above. The subtraction term is defined as the 
squared resonant amplitude with the top quark being on-shell, except for the propagator denominator, which is kept 
as a general Breit-Wigner function 
\begin{equation}
	\left| \mathcal{M}_{2\to 3}^{\tt sub} \right|^2 = 
		\frac{m_t^2 \Gamma_t^2} {(p_t^2-m_t^2)^2 + m_t^2 \Gamma_t^2} 
		\left| \mathcal{M}_{2\to 3}^{\tt res} \right|^2_{p_t^2=m_t^2} .
\end{equation}
When the top quark is exactly on-shell, the subtraction term is equal to the full $2 \to 3$ matrix element, while it decreases as a 
Breit-Wigner distribution when the top quark moves away from its pole. This method has the advantage that the resulting cross section 
retains the non-resonant interferences of the two processes. We have checked that the total cross section after subtraction is 
independent of the top quark width.

Other diagrams with different final states can also include on-shell propagators but for most of them only in very specific 
configurations, e.g., mass degeneracy between $\tilde{t}_1$ and $\tilde{t}_2$ or between $\tilde{t}_1$ and $\tilde{b}_1$. Those cases 
are not relevant for our study of $\tilde{\chi}^0_1 \tilde{t}_1$ co-annihilation.

Another numerical instability arises from the fact that, in case of co-annihilation into quark and photon, also the 
external photon of the real emission subprocess $\tilde{\chi}_n^0 \tilde{q}_i \rightarrow q g \gamma$ may become soft 
in certain regions of phase space, rendering the numerical integration unreliable. This issue can be addressed by 
introducing a cut-off on the photon energy in order to exclude the corresponding part of the phase space. This soft 
behavior (and the associated cut-off dependence) would vanish when including also electroweak corrections, which is, 
however, beyond the scope of this work. Moreover, as we have seen in Sec.~\ref{Sec:Pheno}, the impact of this process 
in the scenarios considered in the present work is negligible.

\subsection{Numerical results}

Let us now discuss in detail the impact of the one-loop corrections on the co-annihilation cross sections 
in our three scenarios of Tab.~\ref{Tab:Scenarios}. We have calculated radiative corrections to two types of processes,
one with a Higgs boson and one with a vector boson in the final state. We have seen that at tree-level the processes with
the Higgs boson final state are dominated by a $t$-channel stop exchange, whereas the processes with a gauge vector boson 
are a mixture of all possible contributions (see Fig.~\ref{Fig:TreeChannels}).
\begin{figure*}[t]
	\includegraphics[scale=0.42]{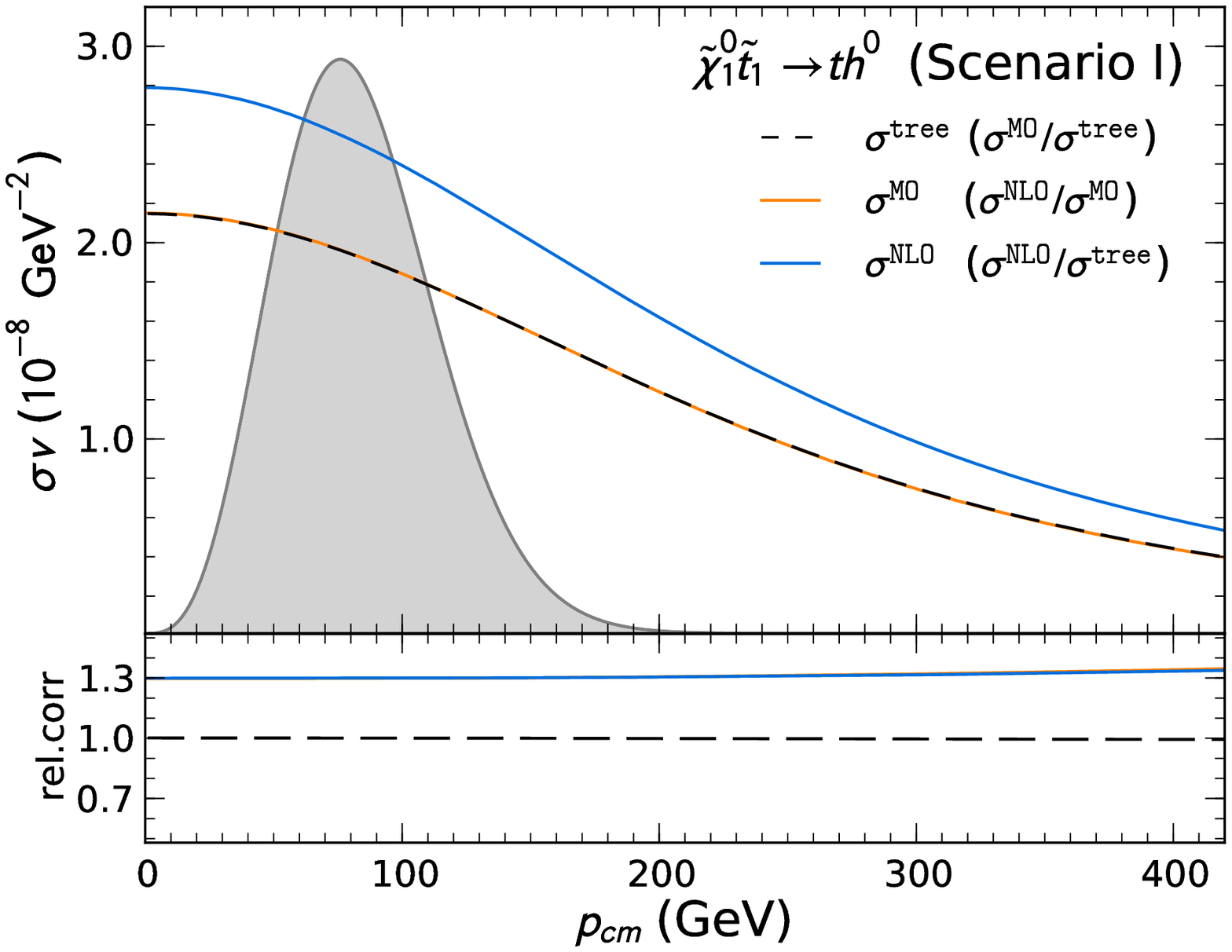}
	\includegraphics[scale=0.42]{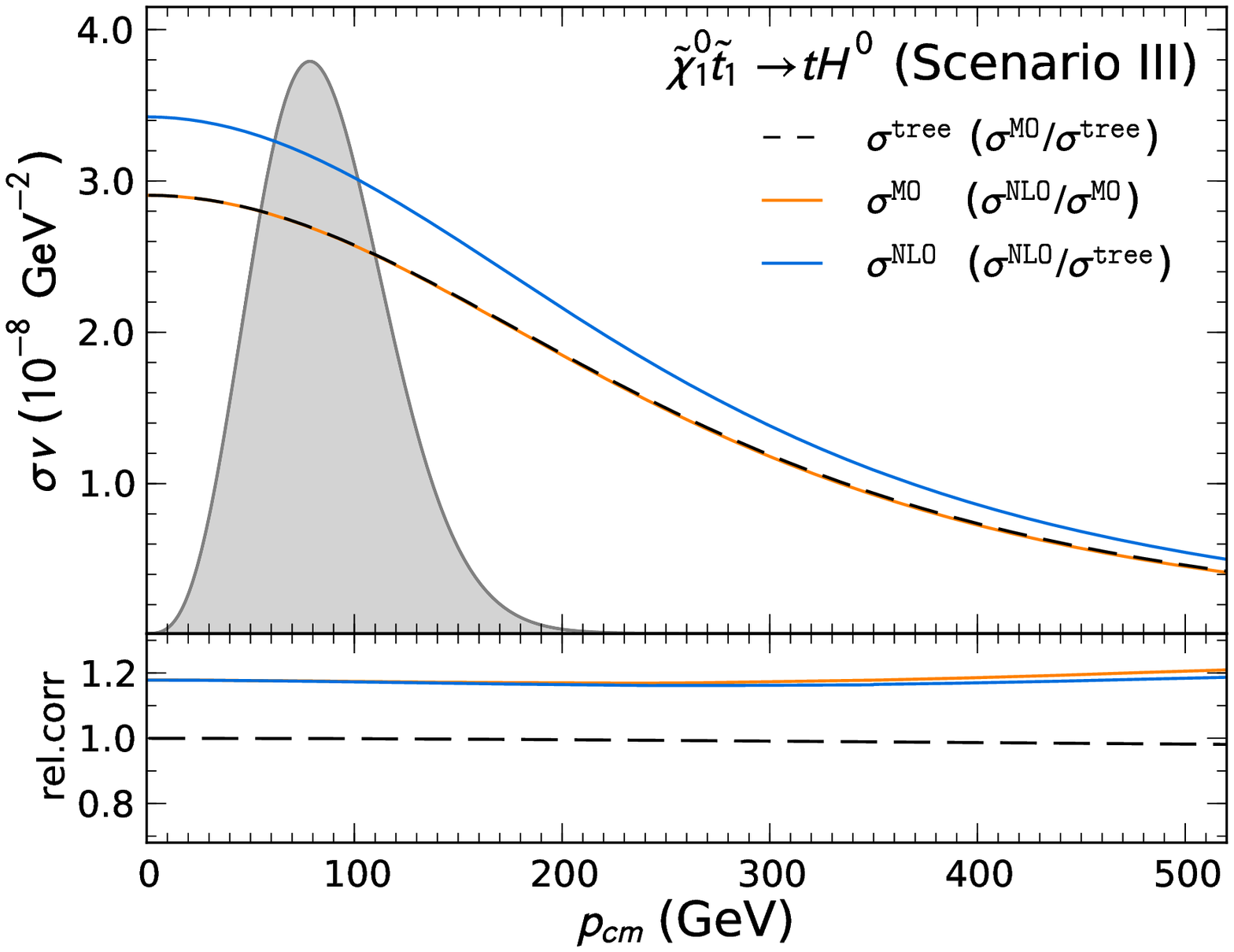} \\
	\includegraphics[scale=0.42]{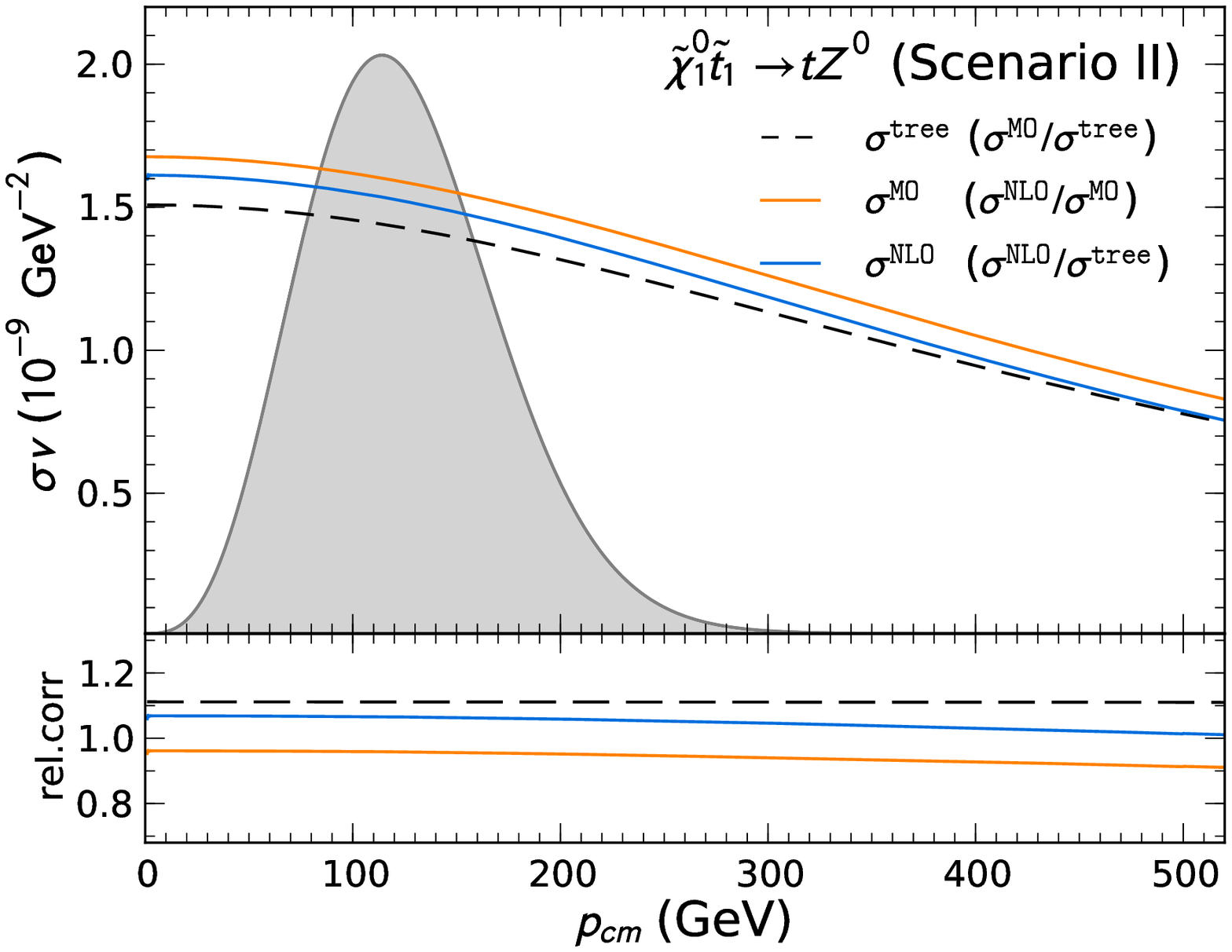}
	\includegraphics[scale=0.42]{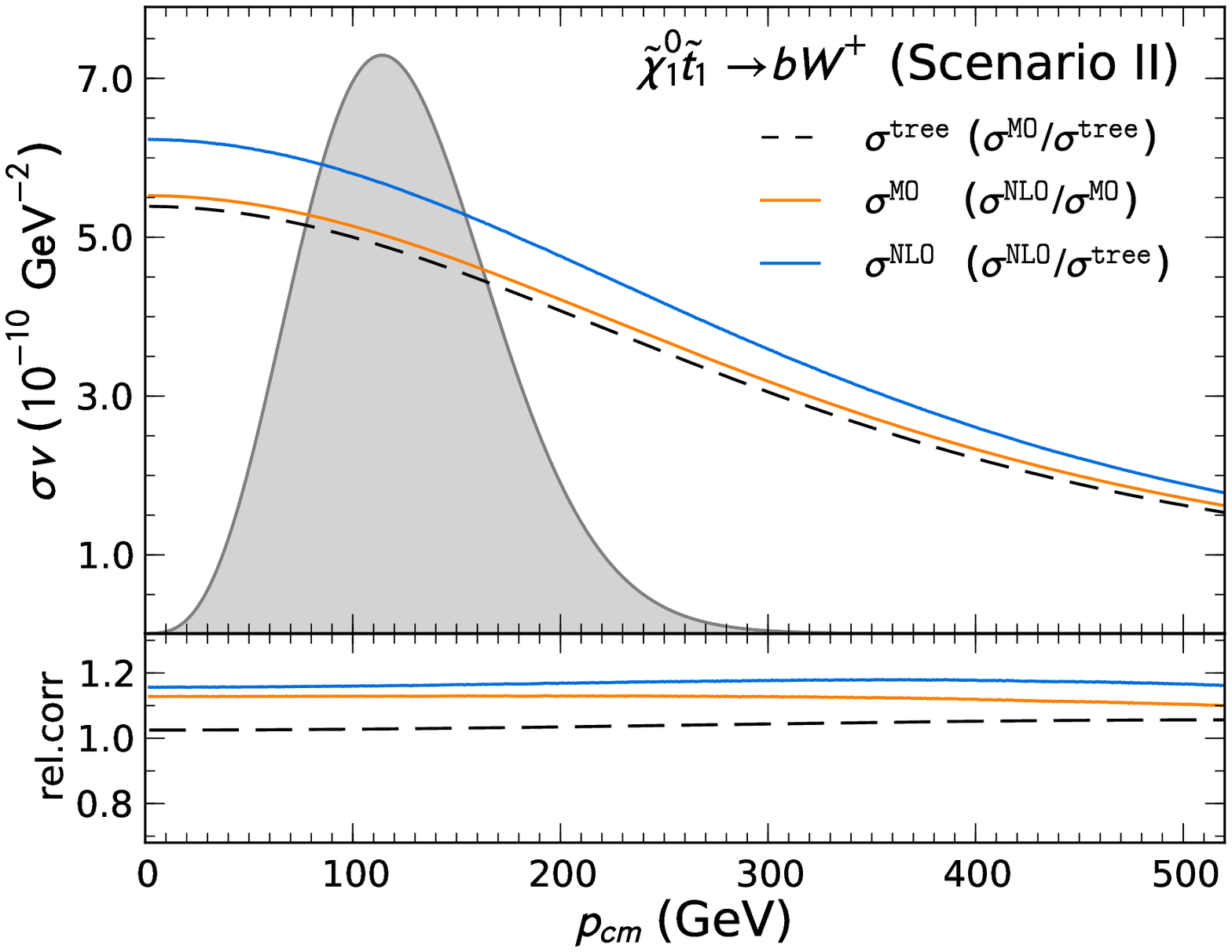}
	\caption{Tree-level (black dashed line), full one-loop (blue solid line) and {\MO} (orange solid line) cross sections 
	for selected co-annihilation channels in the scenarios of Tab.~\ref{Tab:Scenarios}. The upper part of each plot shows 
	the absolute value of $\sigma v$ together with the thermal distribution (in arbitrary units), whereas the lower part 
	shows the corresponding relative shifts (second item in the legend).}
	\label{Fig:CrossSections}
\end{figure*}

These different compositions of the cross sections influence also the impact of various types of loop corrections 
which are displayed in Fig.~\ref{Fig:CorrContrib}. This figure shows a break down of the total 
next-to-leading correction to the cross section $\sigma v$ (without the tree-level contribution) into several UV finite contributions for both types of processes, $\tilde{\chi}_1^0 \tilde{t}_1 \to t h^0$ (scenario I) and 
$\tilde{\chi}_1^0 \tilde{t}_1 \to t Z^0$ (scenario II). 
Even though all contributions are UV finite, the box, vertex and real part of the correction are still IR divergent. This leads to a certain ambiguity in their exact definition. Each contribution contains an uncancelled pole along with an uncancelled logarithm of the large factorization scale. These large logarithms cause the box contribution to be artificially large and drive the real corrections (which in our case is a sum of the soft-gluon part and the hard radiation) to be negative.


Comparing the different loop contributions for the scalar and vector boson final states, one notices that the box 
and propagator corrections in the case of the Higgs boson final states are enhanced. This can be traced back to 
the fact that the cross section 
with a Higgs boson in the final state is dominated by the $t$-channel exchange. One of the loop corrections to the 
$t$-channel entails a correction to the stop propagator and a box diagram where a gluon is exchanged between the final state 
quark and the initial state squark. The enhanced box and propagator corrections lead to a large overall NLO correction 
in the case of the co-annihilation cross section with the Higgs boson.

We show the cross sections of the respectively most relevant channel in each scenario 
in Fig.~\ref{Fig:CrossSections} and compare our tree-level calculation, the effective tree-level calculation implemented 
in {\MO} and our full one-loop calculation. 
The upper parts show the cross sections $\sigma v$, while the lower panels show the ratio between 
the different cross sections.

For scenario I, where we show the channel $\tilde{\chi}_1^0 \tilde{t}_1 \to t h^0$, we have numerical agreement between 
our tree-level and the {\MO} calculation. The one-loop contributions increase the cross section by about 30\%
caused by the large contribution from the box diagrams and propagator corrections as discussed above. 
We observe a similar behavior for scenario III, where the final state with a heavy Higgs boson $H^0$ is dominant. 
Here, the one-loop cross section lies about 18 -- 20\% above the tree-levels, which again agree well among each other. 

\begin{figure*}[th]
	\includegraphics[scale=0.42]{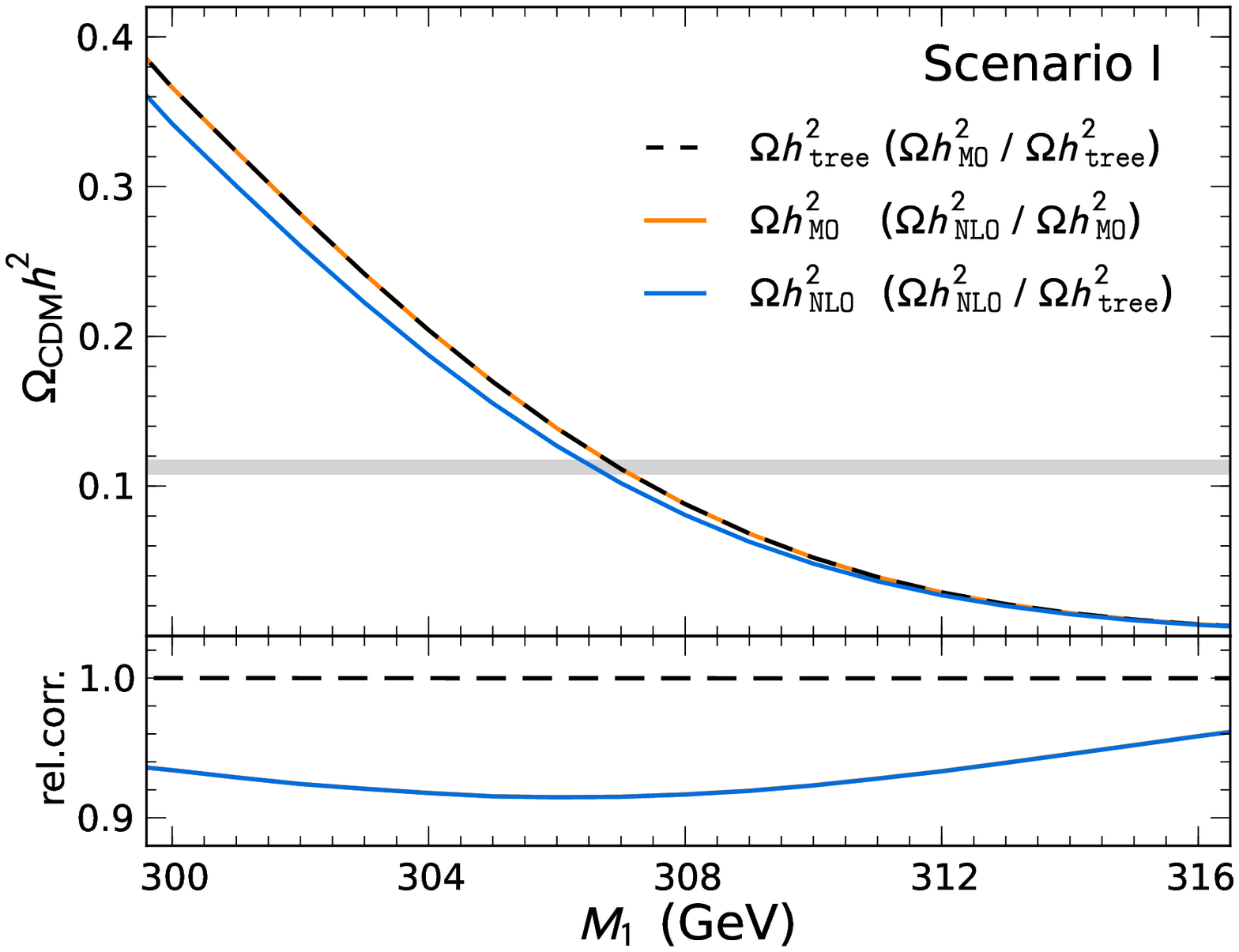}
	\includegraphics[scale=0.42]{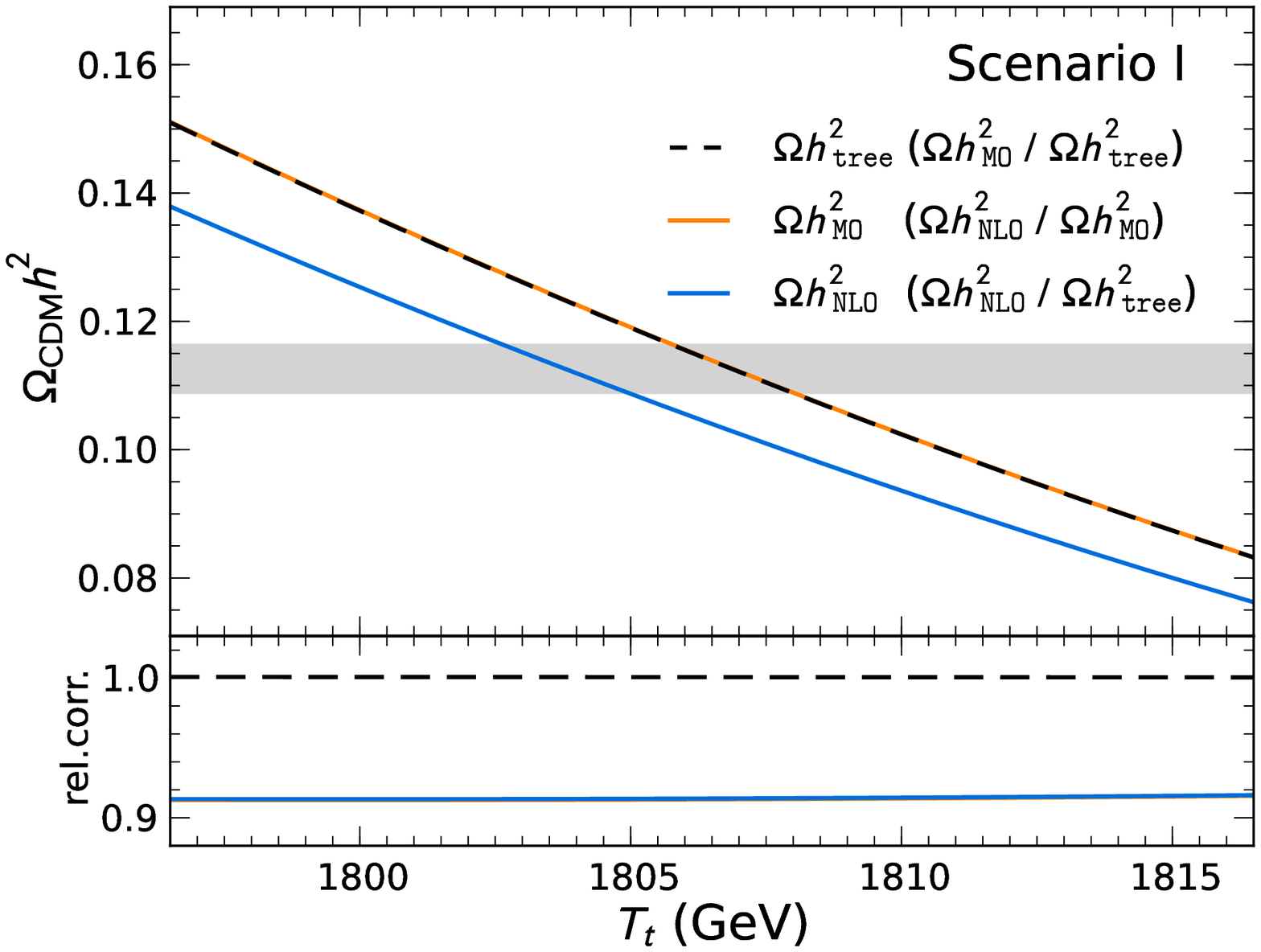}\\
	\caption{The neutralino relic density $\Omega_{\chi}h^2$ as a function of $M_1$ 
	(left) and $T_t$ (right) in our scenario I calculated using different co-annihilation cross sections: 
	default {\MO} (orange solid line), tree-level (black dashed line), and full one-loop (blue solid line). The gray band indicates the 
	favored range according to Eq.\ (\ref{Eq:WMAP}). The lower part shows the relative impact 
	of the one-loop correction on the relic-density compared to the tree-level calculation (second item in the legend).}
	\label{Fig:Relic1DScenario1}
\end{figure*}

In case of co-annihilation into a quark and an electroweak gauge boson, there is a few percent difference between our 
tree-level and the one provided by {\MO}. This difference stems from the fact that both tree-levels use different parameters.
Our tree-level uses input parameters defined through the renormalization scheme discussed in detail in Sec.~\ref{Sec:RenScheme}.
It differs in several points from the parameters used by {\MO}. More precisely, the shift between the two tree-levels is largely 
due to a different definition of the squark mixing angles, which enter the calculation through the different interactions between 
squarks and quarks, e.g., the neutralino-squark-quark vertex. 

The different influence of various definitions of the mixing angle on the two classes of processes we have calculated can be understood 
as follows: In the case of the Higgs boson final state, which is dominated by a squark-exchange in the $t$-channel, the mixing angle 
$\theta_{\tilde{t}}$ enters the squark-squark-Higgs and the neutralino-squark-quark vertices. The internal propagator has to be
summed over the two possible squark mass eigenstates, $\tilde{t}_1$ and $\tilde{t}_2$, making the result less sensitive to the 
exact value of the mixing angle. For the $s$-channel dominated co-annihilation into $t Z^0$ or $b W^+$, the situation is quite 
different. Here, the mixing angle appears in a single neutralino-squark-quark vertex, where the external squark is ``fixed'' to be 
$\tilde{t}_1$. The corresponding matrix element is therefore rather sensitive to changes in the mixing angle, which explains the
observed difference between the two tree-level curves.

\section{Impact on the neutralino relic density \label{Sec:Results}}

The main purpose of this analysis is to investigate the impact of higher order corrections
on the neutralino relic density.

Our numerical implementation of the calculation described in Sec.~\ref{Sec:Calculation} 
is used as an extension to the public package {\MO} in order to evaluate the effect of the 
one-loop corrections on the neutralino relic density. We stress that our implementation is 
general so that it can be used for any neutralino-sfermion co-annihilation process, even if 
we focus in this study on the case of $\tilde{\chi}^0_1 \tilde{t}_1$, which is the most 
relevant process of this kind within the MSSM. Our numerical code is linked to {\MO} in such 
a way that all relevant parameters, i.e.\ the masses and mixings of all particles, are 
passed between the two codes in a consistent way. In particular, we use {\SPheno} to 
compute the supersymmetric mass spectrum for our characteristic scenarios as described in Sec.~\ref{Sec:Pheno}.

In this section we compare the neutralino relic density obtained from the three different 
cross section calculations, which have been described in Sec.~\ref{Sec:Calculation}: the one 
used by default in {\MO}, evaluated by {\CHep} \cite{CalcHEP} at tree-level, 
our cross section at tree-level, and our calculation including the full next-to-leading order SUSY-QCD corrections. 
The impact of the corrections compared to the tree-level results is studied 
for the three scenarios defined in Tab.~\ref{Tab:Scenarios}.

\begin{figure*}[th]
	\includegraphics[scale=0.42]{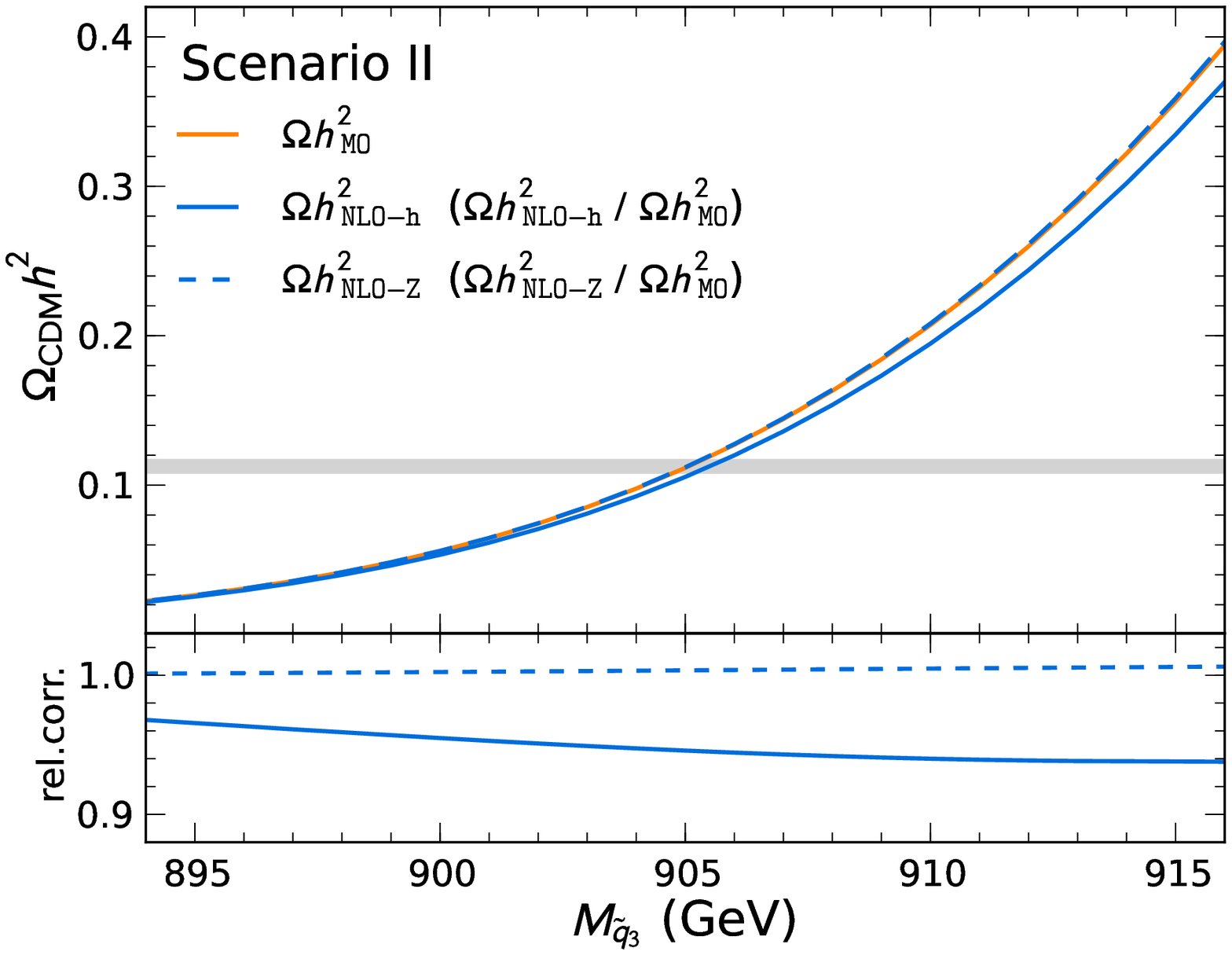}
	\includegraphics[scale=0.42]{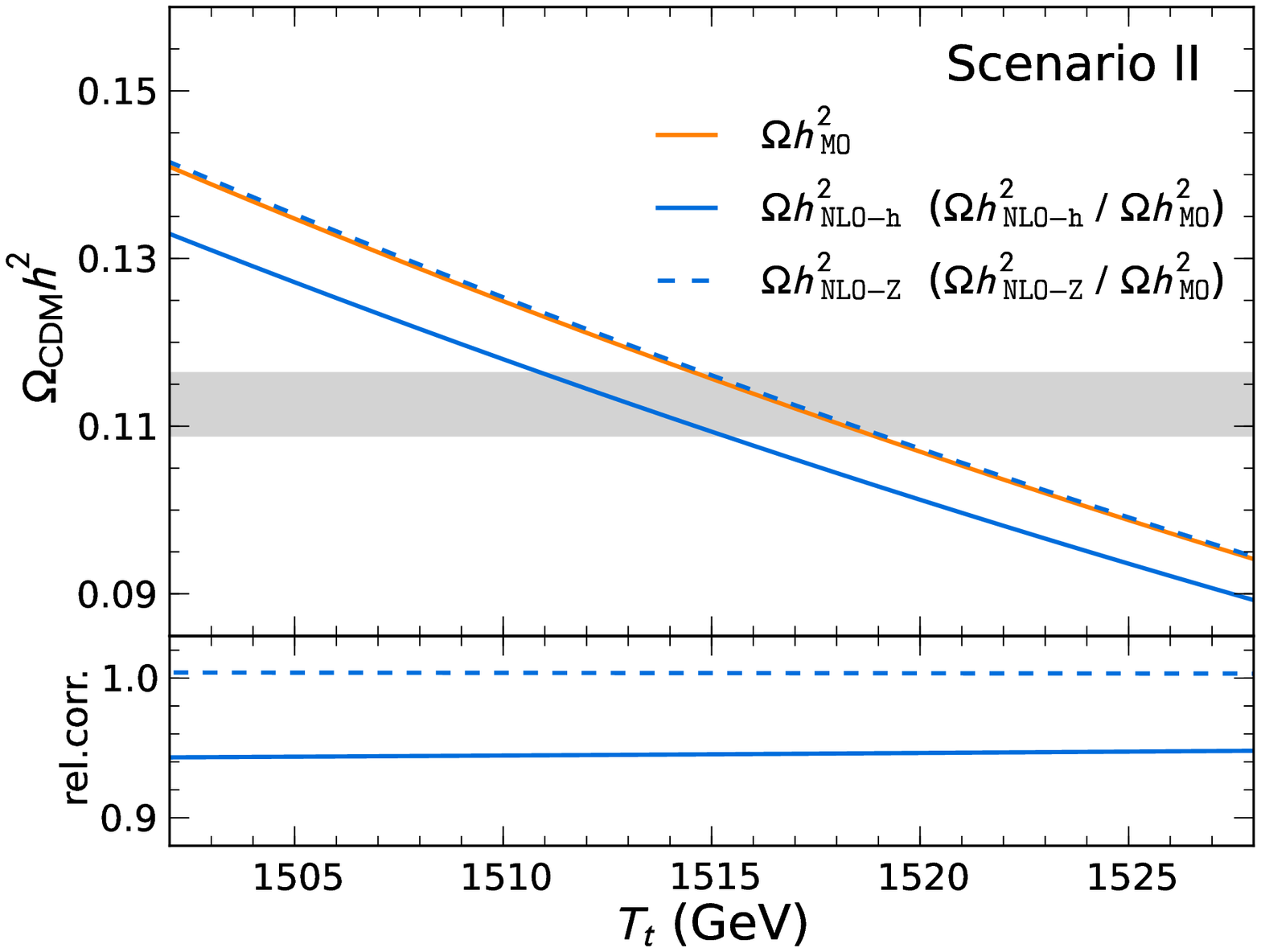}\\
	\caption{The neutralino relic density $\Omega_{\chi}h^2$ as a function of $M_{\tilde{q}_3}$ 
	(left) and $T_t$ (right) in our scenario II calculated using different co-annihilation cross sections: 
	default {\MO} (orange solid line), one-loop correction only for the $t h^0$ final state (blue solid line), 
	and one-loop correction only for the $t Z^0$ final state (blue dashed line). The gray band indicates the 
	favored range according to Eq.\ (\ref{Eq:WMAP}). The lower part of the figure shows the relative impact 
	of the one-loop correction on the relic-density compared to {\MO} (second item in the legend).}
	\label{Fig:Relic1DScenario2}
\end{figure*}

First, we focus on scenario I. We study the change of the relic density when a single input parameter is varied
around our scenario I. In Fig.~\ref{Fig:Relic1DScenario1}, we show $\Omega_{\chi}h^2$ as a function of the 
bino mass parameter $M_1$ and the trilinear coupling parameter $T_t$, calculated on the basis of the aforementioned 
three calculations for the neutralino-stop co-annihilation. It is clearly visible that the relic density is 
very sensitive to variations of the bino mass parameter. For higher values of $M_1$ the predicted relic density 
decreases rapidly due to a smaller mass splitting between the lightest neutralino and the lightest stop, which 
enhances the neutralino-stop co-annihilation and in addition the stop-stop annihilation. In contrast, slightly 
lower values for the bino mass parameter increase the mass difference and suppress the contribution of co-annihilation 
processes in favor of neutralino-neutralino annihilation. The predicted relic density is then higher due to the 
absence of co-annihilation. Within the area which is favored by the measurements of 
WMAP, where the studied neutralino-stop coannihilation is dominant, a clear shift of the predicted relic density is 
visible when going from the default value calculated by {\MO} to the one calculated using our full next-to-leading 
order result.

The impact of the presented SUSY-QCD corrections to the given neutralino-stop co-annihilation processes is even 
better visible in the lower part of Fig.~\ref{Fig:Relic1DScenario1}, where we show the relative correction, i.e.\ the 
ratio of the relic density calculated with our full one-loop co-annihilation cross section to the one included by default 
in {\MO} and our tree-level, respectively. 
For scenario I, our calculations result in a relative correction of about 9\%. This can be explained by the lightest Higgs 
final state, which has a contribution of around 38.5\% to the total (co-)annihilation cross section with a corresponding correction of around 30\% (see Fig.~\ref{Fig:CrossSections}). With the 
current experimental uncertainty of about 3\% according to Eq.~(\ref{Eq:WMAP}), the impact of the presented 
corrections is significant and thus important to be taken into account.

The relic density is less sensitive to varying the trilinear coupling parameter $T_t$ around the value in 
scenario I ($T_t=1806.5$ GeV). This is depicted on the right-hand side of Fig.~\ref{Fig:Relic1DScenario1}. 
Here, the difference between the uncorrected and corrected relic density in the cosmologically favored 
region corresponds to a difference of 3 GeV in the parameter $T_t$.

\begin{figure*}
	\includegraphics[scale=0.46]{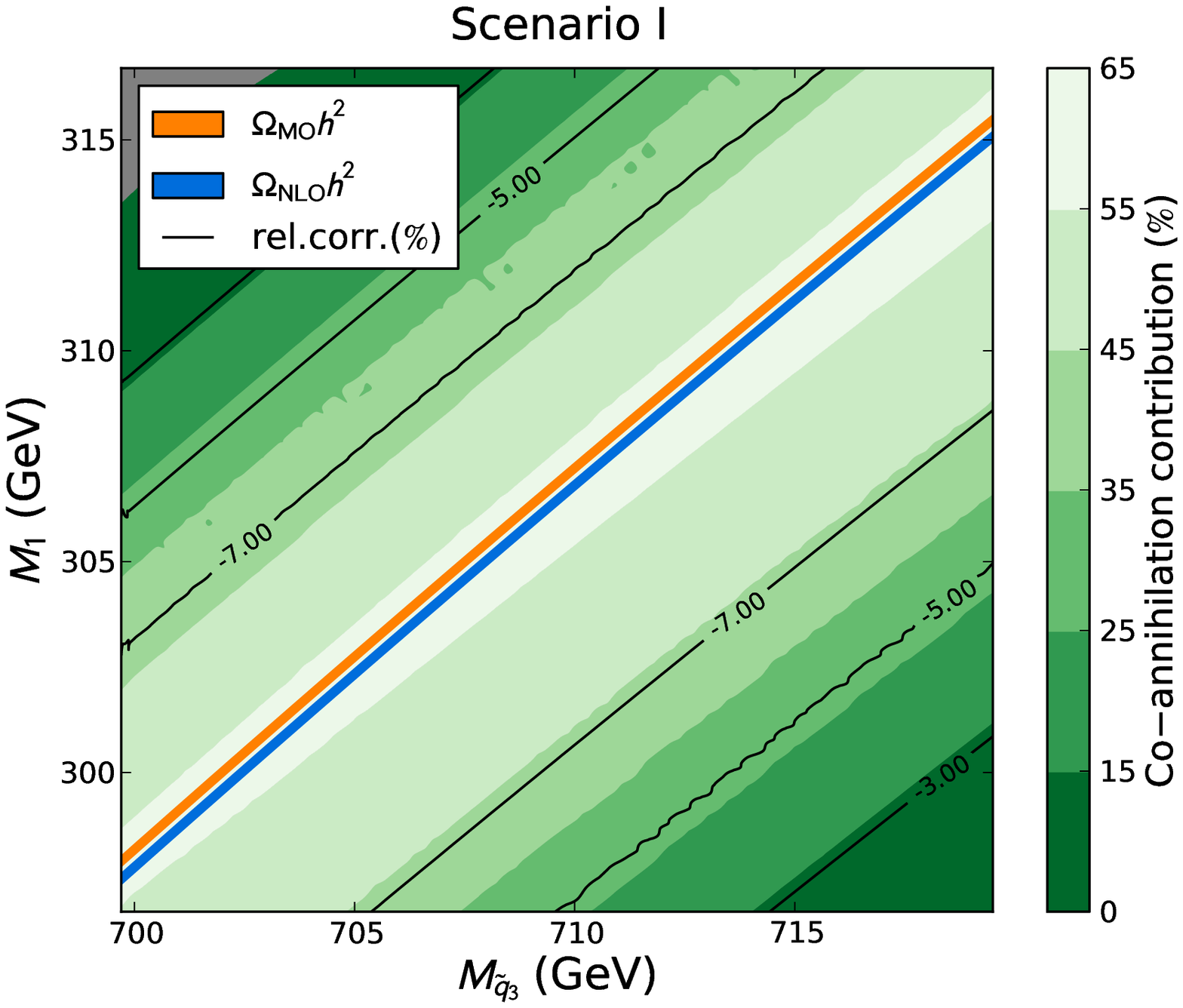}
	\includegraphics[scale=0.46]{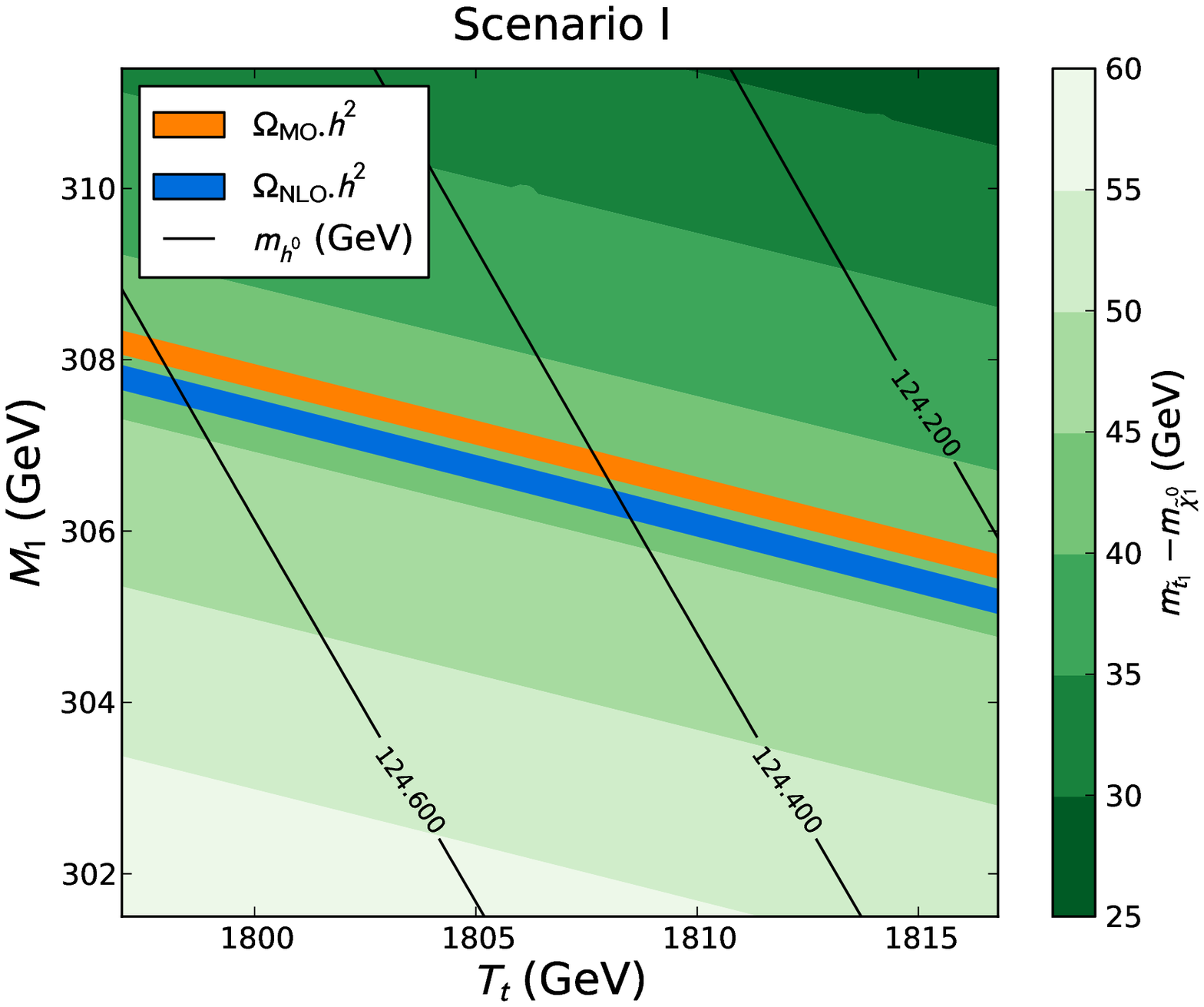}\\
	\includegraphics[scale=0.46]{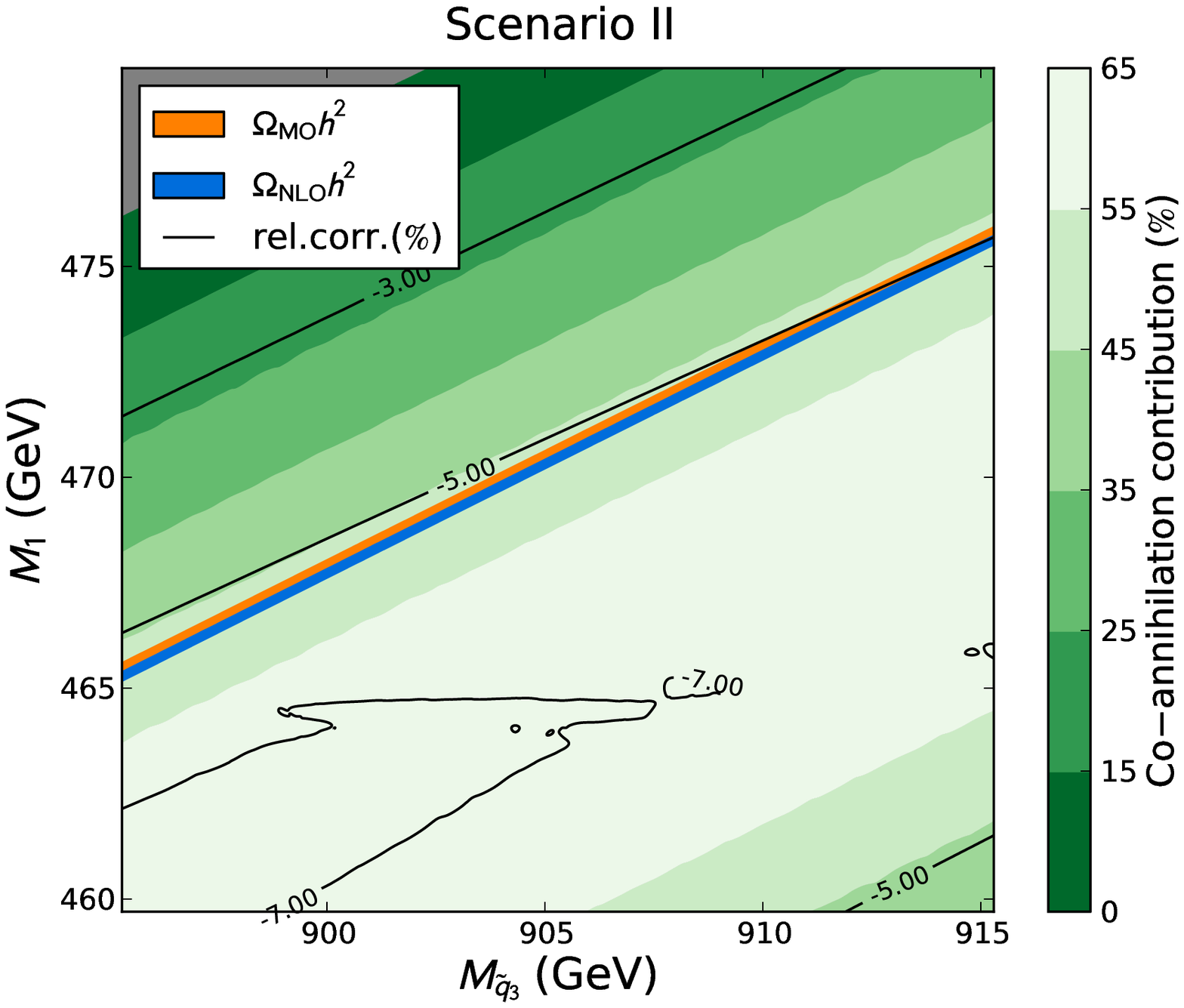}
	\includegraphics[scale=0.46]{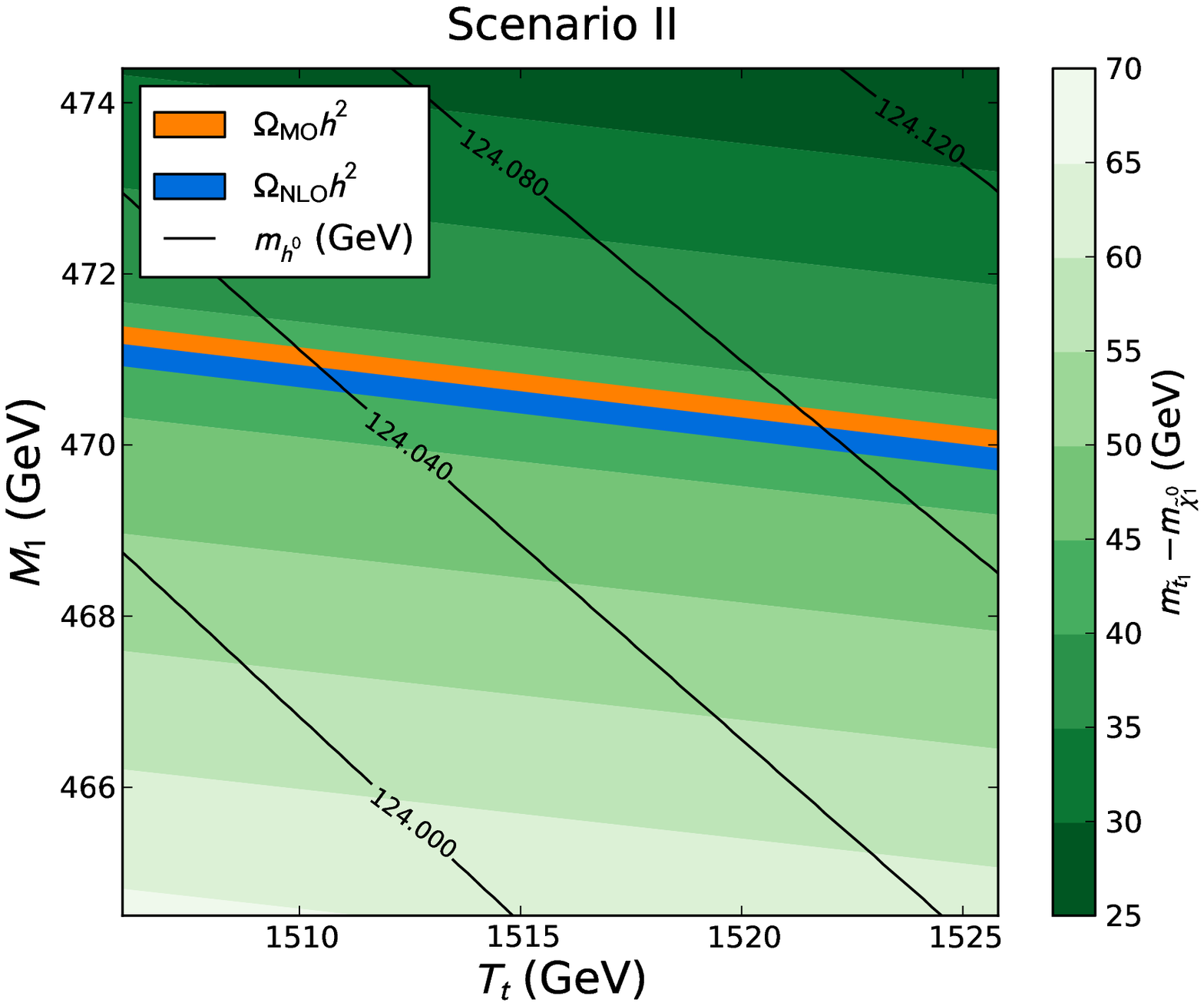}\\
	\includegraphics[scale=0.46]{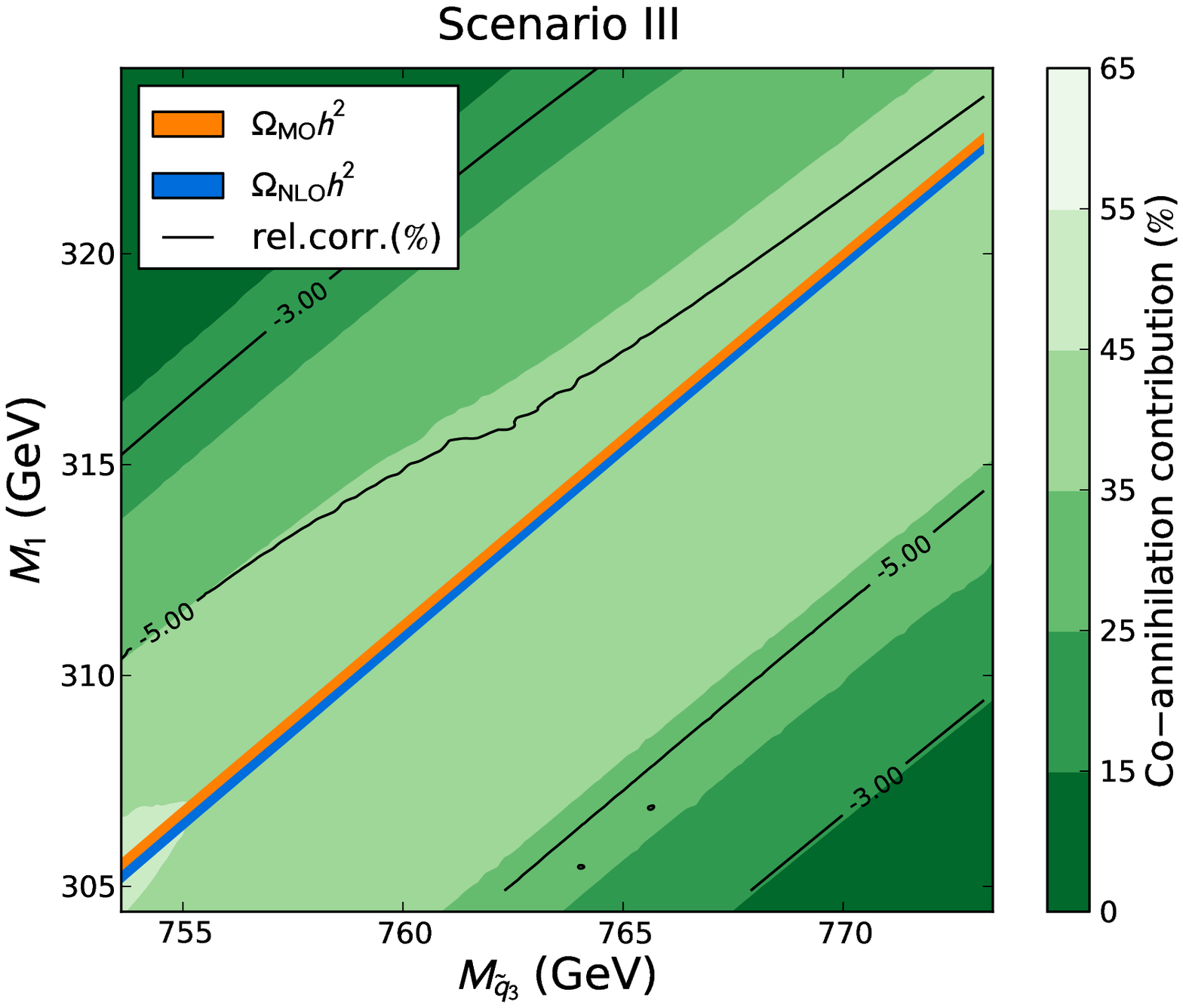}
	\includegraphics[scale=0.46]{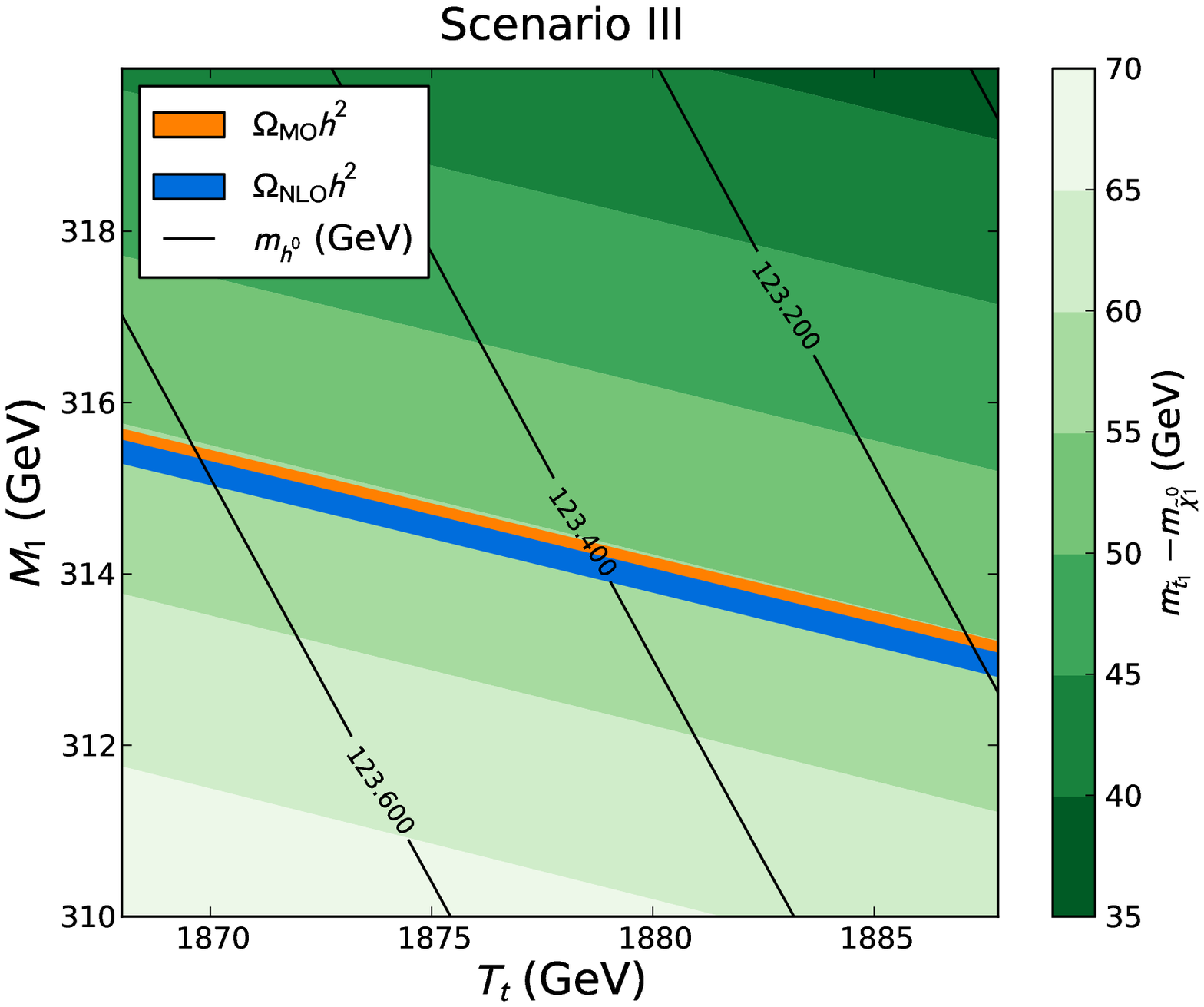}
	\caption{WMAP-compatible relic density bands from the default {\MO} calculation (orange) and our one-loop calculation for co-annihilation (blue) in the $(M_{\tilde{q}_3},M_1)$ (left) and $(T_t,M_1)$ (right) planes. In the plots on the left hand side the relative contribution of co-annihilation processes is shown in green contour, and the relative impact of the one-loop corrections on the relic density in black lines. The plots on the right hand side show the LSP-NLSP mass difference in green contour, and the lightest neutral Higgs boson mass in black lines.}
	\label{Fig:Relic2D}
\end{figure*}

One can infer more about the impact of the full next-to-leading order corrections in scenario I when
looking at the first row of Fig.~\ref{Fig:Relic2D}. On the left, the WMAP favored region is shown as a 
function of two parameters - the mass parameter of the third generation of squarks $M_{\tilde{q}_3}$ 
and the bino mass parameter $M_1$. In the same plot solid black contour lines denote the relative 
impact of our correction to the default {\MO} relic density. As the co-annihilation into the lightest 
Higgs is the dominant contribution to the total (co-)annihilation cross section around the WMAP-favored 
region in this scenario, and as it receives large corrections, a relative correction 
of up to 9\% on the relic density is observed. The correction is larger than current experimental 
uncertainties, which results in two separated WMAP-favored $1 \sigma$-bands corresponding to the default {\MO} 
calculation (orange) and our full one-loop SUSY-QCD calculation (blue).

The cosmologically allowed band follows a straight line in the $M_1$-$M_{\tilde{q}_3}$ plane corresponding 
to a constant mass difference between the lightest neutralino and the lightest stop of about 40 GeV.
Above this band where the neutralino becomes heavier and the mass difference decreases, the stop-stop 
annihilation becomes dominant. As it has typically a significant higher cross section than the co-annihilation, 
it leads to a relic density which is too small. For large values of $M_1$ (in the gray area in the upper left 
corner) the stop becomes the lightest supersymmetric particle, which is disfavored as a suitable dark matter 
candidate both for its electric and color charge.

In the opposite direction, below the allowed band, the neutralino-stop and stop-stop (co-)annihilation are 
Boltzmann suppressed by a larger mass difference and neutralino annihilation becomes dominant. However, it has a 
lower cross section, such that the relic density becomes too big. 

To conclude our analysis of scenario I, on the right plot in Fig.~\ref{Fig:Relic2D} we show WMAP preferred regions in 
the $(T_t,M_1)$ plane. Again, a clear separation of the two bands is visible, together with the small 
dependence on the trilinear coupling parameter $T_t$ (as already discussed for Fig.~\ref{Fig:Relic1DScenario1}). 
In different green colors, the mass difference between the lightest and next-to-lightest supersymmetric particle is 
depicted supporting the claim that the cosmologically favored region follows a contour of a constant mass difference 
around $40-45~{\rm GeV}$. 
The solid black lines show the mass of the lightest Higgs boson in the selected parameter space. 
One can see that the whole WMAP favored region in this plot lies within the recent Higgs mass limit 
$125.2~{\rm GeV} \pm 0.9~{\rm GeV}$ as reported in \cite{ATLAS2012update}. It is also interesting to note
that the cosmological constraints from WMAP are at the moment more stringent than the current bounds on a 
Higgs-like particle.

Let us now focus on scenario II, which differs in several crucial features from the previously analyzed scenario I.
One example is that the total co-annihilation cross section has two dominating contributions from co-annihilation 
into the lightest Higgs and into the $Z$-boson.
In Fig.~\ref{Fig:Relic1DScenario2} we show separately the effect of SUSY-QCD corrections to each of the two dominant 
processes as a function of the parameters $M_{\tilde{q}_3}$ and $T_t$. One can see distinctly different effects higher 
order corrections have on each process. As in scenario I, large corrections to co-annihilation into the lightest Higgs 
bosons lead to a change of up to 6\% in the relic density even though its relative importance in the total cross section 
dropped to 24\% compared to scenario I. On the other hand corrections to co-annihilation into the $Z$-boson are small 
(see Fig.~\ref{Fig:CrossSections}) and also differ in sign. This leads to a reduction of the impact of SUSY-QCD corrections 
on the relic density in scenario II. The consequences can be seen in the second row of Fig.~\ref{Fig:Relic2D}. 
One sees that due to the smaller correction of about 5-6\%, the two bands corresponding to the original {\MO} relic density 
(orange) and the one obtained including our SUSY-QCD corrections (blue) overlap.

Scenario II is different from the others also in that the preferred WMAP region lies outside of the area with maximal 
co-annihilation fraction. This is a direct consequence of the importance of the co-annihilation into the $Z$-boson which 
has a smaller cross section and so in total co-annihilation is not efficient enough to bring the relic density 
down to the level measured by WMAP (the allowed region receives sizable contributions from the stop annihilations).

In contrast to other scenarios, in scenario II co-annihilation dominate in a region where the mass difference between the stop and 
the lightest neutralino is larger (about $70~{\rm GeV}$). This can be traced back to the masses of the lightest 
neutralino and the stop, which are much heavier than in the other two scenarios. As a result the freeze-out temperature, 
which is proportional to the mass of the dark matter particle, is higher. This means that the same Boltzmann suppression 
which for scenario I was obtained for a mass difference $40-45~{\rm GeV}$, is now reached for a larger mass splitting of 
$70~{\rm GeV}$.
 
In the third scenario, the light CP-even Higgs boson is the dominant contribution to neutralino-stop co-annihilation and 
the characteristics of the plots in Fig.~\ref{Fig:Relic2D} are similar to scenario I. As the correction to the top-$H^0$ 
final states is not as large as for the top-$h^0$ final state in this example point (see Fig.~\ref{Fig:CrossSections}), 
the overall impact on the relic density is thus smaller than for scenario I. A relative correction between $5 \%$ to $6 \%$ 
is reached. Nevertheless, a shift from the WMAP favored region calculated by {\MO} to the one calculated with the 
one-loop SUSY-QCD corrections is visible.
An interesting feature can be observed by comparing the plots in the second column of Fig.~\ref{Fig:Relic2D} regarding 
the Higgs mass. Whereas for scenario I and III the Higgs mass is decreasing with an increasing trilinear coupling parameter, 
it is the opposite for scenario II. Analyzing Eq.\ (\ref{Eq:HiggsMass}), where the maximal contribution is obtained from 
a stop mixing for $|X_t| \sim \sqrt{6} M_{\rm SUSY}$, this effect becomes clear. 
In scenario II, we find $|X_t| < \sqrt{6} M_{\rm SUSY}$ and the Higgs mass grows with increasing $X_t$, whereas in the other scenarios 
$|X_t| > \sqrt{6} M_{\rm SUSY}$ and the Higgs mass decreases as $X_t$ gets larger.
In addition, it is interesting that in comparison to the other two scenarios, the preferred region lies in the band where the mass splitting between the neutralino and stop is already around $55-60~{\rm GeV}$.

Studying the three different characteristic scenarios, we saw that the impact of the one-loop corrections on the predicted relic density of dark matter can be more important than the current experimental uncertainty by the WMAP observations. Therefore it is necessary to take them into account for a theoretical prediction of the neutralino relic density.

\section{Conclusions \label{Sec:Conclusions}}

One of the relevant mechanisms to obtain the observed relic density of dark matter relies on the presence of co-annihilation of the dark matter candidate with another particle which is almost degenerate in mass. We have studied this situation within the Minimal Supersymmetric Standard Model (MSSM), where the dark matter candidate is the lightest of the four neutralinos. More precisely, we have focused on the case of co-annihilation with a relatively light stop.

We have demonstrated that the interpretation of a new boson with a mass of about 126 GeV in terms of the lightest Higgs boson within the MSSM favors this situation due to the necessity of an important mass splitting in the stop sector. This results in general in one relatively small mass eigenvalue. If this value is close enough to the neutralino mass, co-annihilations are the dominant annihilation channel driving the Boltzmann equation. The important stop mass splitting is mostly realized if the trilinear coupling parameter $T_t$ in the stop sector is sizable. This in turn increases the relative importance of the neutralino-stop co-annihilation into a top quark and a Higgs boson, which is driven by precisely the same trilinear coupling. Other channels, such as co-annihilation into a top (bottom) quark and a $Z$ ($W$)-boson are present, but mostly subdominant.

In order to keep up with the current and future experimental accuracies, a reduction of the theoretical uncertainty is necessary. The main source of uncertainty on the particle physics side comes from the calculation of the (co-)annihilation cross section, which governs the Boltzmann equation and thus the prediction of the dark matter relic density. To this end, we have calculated the co-annihilation of a neutralino with a stop into final states containing electroweak gauge or Higgs bosons at one-loop order in SUSY-QCD. In particular, we have defined a renormalization scheme, which can consistently be applied to all neutralino annihilation and co-annihilation processes. Infrared singularities are handled using the phase-space slicing method. The present work is complementary to previous publications on radiative corrections to neutralino pair-annihilation \cite{DMNLO_AFunnel, DMNLO_mSUGRA, DMNLO_NUHM} or co-annihilation with a stop into a top quark and a gluon or a bottom quark and a $W$-boson \cite{Freitas2007}. 
In order to obtain a consistent implementation of all co-annihilation processes, including the missing case of a gluon final state will be necessary. This step is, however, postponed to a later publication.

In summary, the impact of the one-loop corrections on the predicted relic density of dark matter can be more important than the current experimental uncertainty by the WMAP observations. The presented corrections are therefore essential in predicting the neutralino relic density for a given parameter point or when extracting SUSY parameters from cosmological measurements. This will become even more important when better limits will be derived from the data of the Planck satellite in a very near future. 

\acknowledgements

The authors would like to thank A.~Freitas, M.~Meinecke and C.~Yaguna for helpful
discussions, W.~Porod for communication regarding the {\SPheno} code, and A.~Pukhov
for providing us with the necessary functions to implement our results into the
{\MO} code. This work is supported by DAAD/EGIDE, Project No.\ PROCOPE 54366394. 
The work of J.H.\ and B.H.\ (in part) is supported by the
Landesexzellenz-Initiative Hamburg ``Connecting particles to the cosmos''. J.H.\
acknowledges support from the CMIRA program of the R\'egion Rh\^one-Alpes and would like to thank the LAPTh for its hospitality during her stays. The work of
M.K.\ is supported by the Helmholtz Alliance for Astroparticle Physics. The work
of Q.L.B.\ is supported by a Ph.D.\ grant of the French Ministry for Education and
Research. Q.L.B.\ would like to thank the theory group of IPN Lyon for its hospitality.

\appendix

\bibliographystyle{apsrev}

\begin{thebibliography}{00}

\bibitem{WMAP7}
  The WMAP collaboration, E.~Komatsu {\it et al.},
  Astrophys.\ J.\ Suppl.\ {\bf 192} (2011) 18
  [arXiv:1001.4538 [astro-ph.CO]].  

\bibitem{GondoloGelmini}
  P.~Gondolo and G.~Gelmini,
  Nucl.\ Phys.\ B {\bf 360} (1991) 145.

  \bibitem{GriestSeckel}
    K.~Griest and D.~Seckel,
    Phys.\ Rev.\  D {\bf 43} (1991) 3191.

\bibitem{EdsjoGondolo}
    J.~Edsjo and P.~Gondolo,
    Phys.\ Rev.\  D {\bf 56} (1997) 1879
    [arXiv:hep-ph/9704361].

\bibitem{Hamann}
  J.~Hamann, S.~Hannestad, M.~S.~Sloth and Y.~Y.~Y.~Wong,
  Phys.\ Rev.\ D {\bf 75} (2007) 023522
  [arXiv:astro-ph/0611582].

\bibitem{Arbey}
  A.~Arbey and F.~Mahmoudi,
  Phys.\ Lett.\ B {\bf 669} (2008) 46
  [arXiv:0803.0741 [hep-ph]].

  \bibitem{SPheno}
  W.~Porod,
  Comput.\ Phys.\ Commun.\  {\bf 153} (2003) 275
  [arXiv:hep-ph/0301101]; \\
  W.~Porod and F.~Staub,
  Comput.\ Phys.\ Commun.\  {\bf 183} (2012) 2458
  [arXiv:1104.1573 [hep-ph]].

\bibitem{Belanger}
  G.~B\'elanger, S.~Kraml and A.~Pukhov,
  Phys.\ Rev.\ D {\bf 72} (2005) 015003
  [arXiv:hep-ph/0502079].

\bibitem{DarkSusy}
  P.~Gondolo, J.~Edsjo, P.~Ullio, L.~Bergstrom, M.~Schelke and E.~A.~Baltz,
  JCAP {\bf 0407} (2004) 008
  [arXiv:astro-ph/0406204];\\
  P. Gondolo, J. Edsjo, P. Ullio, L. Bergstrom, M. Schelke, E.A. Baltz, T. Bringmann and G. Duda, http://www.darksusy.org

\bibitem{micrOMEGAs2007}
  G.~B\'elanger, F.~Boudjema, A.~Pukhov and A.~Semenov,
  Comput.\ Phys.\ Commun.\  {\bf 177} (2007) 894; \\
  G.~B\'elanger, F.~Boudjema, A.~Pukhov and A.~Semenov,
  Comput.\ Phys.\ Commun.\  {\bf 149} (2002) 103
  [arXiv:hep-ph/0112278].


\bibitem{DMNLO_AFunnel}
  B.~Herrmann and M.~Klasen,
  Phys.\ Rev.\ D {\bf 76} (2007) 117704
  [arXiv:0709.0043 [hep-ph]].

\bibitem{DMNLO_mSUGRA}
  B.~Herrmann, M.~Klasen and K.~Kovarik,
  Phys.\ Rev.\ D {\bf 79} (2009) 061701
  [arXiv:0901.0481 [hep-ph]].

\bibitem{DMNLO_NUHM}
  B.~Herrmann, M.~Klasen and K.~Kovarik,
  Phys.\ Rev.\ D {\bf 80} (2009) 085025
  [arXiv:0907.0030 [hep-ph]].



\bibitem{Sloops2007}
  N.~Baro, F.~Boudjema and A.~Semenov,
  Phys.\ Lett.\ B {\bf 660} (2008) 550
  [arXiv:0710.1821 [hep-ph]].

\bibitem{Sloops2009}
  N.~Baro, G.~Chalons and S.~Hao,
  AIP Conf.\ Proc.\  {\bf 1200} (2010) 1067
  [arXiv:0909.3263 [hep-ph]].

\bibitem{Sloops2010}
  N.~Baro, F.~Boudjema, G.~Chalons and S.~Hao,
  Phys.\ Rev.\ D {\bf 81} (2010) 015005
  [arXiv:0910.3293 [hep-ph]].

\bibitem{Sloops2011}
  F.~Boudjema, G.~Drieu La Rochelle and S.~Kulkarni,
  Phys.\ Rev.\ D {\bf 84} (2011) 116001
  [arXiv:1108.4291 [hep-ph]].

\bibitem{EffCouplings}
  A.~Chatterjee, M.~Drees and S.~Kulkarni,
  arXiv:1209.2328 [hep-ph].

\bibitem{Freitas2007}
  A.~Freitas,
  Phys.\ Lett.\ B {\bf 652} (2007) 280
  [arXiv:0705.4027 [hep-ph]].


\bibitem{StopCoann2}
  	C.~Boehm, A.~Djouadi and M.~Drees,
  	Phys.\ Rev.\ D {\bf 62} (2000) 035012
  	[hep-ph/9911496].

\bibitem{StopCoann1}
    J.~Ellis, K.~A.~Olive and Y.~Santoso,
    Astropart.\ Phys.\ {\bf 18} (2003) 395
    [arXiv:hep-ph/0112113].


\bibitem{EWBG}
      D.~Delepine, J.-M.~G\'erard, R.~Gonzalez Felipe and J.~Weyers,
      Phys.\ Lett.\ B {\bf 386} (1996) 183
      [arXiv:hep-ph/9604440].

\bibitem{naturalSUSY1}
	  M.~Papucci, J.~T.~Ruderman and A.~Weiler,
  		JHEP {\bf 1209} (2012) 035
  		[arXiv:1110.6926 [hep-ph]].

\bibitem{naturalSUSY2}
      R.~Auzzi, A.~Giveon, S.~B.~Gudnason and T.~Shacham,
      arXiv:1208.6263 [hep-ph].

\bibitem{StopCoannLHC1}
      Z.-H.~Yu, X.-J.~Bi, Q.-S.~Yan, P.-F.Yin,
      arXiv:1211.2997 [hep-ph].
	
\bibitem{StopCoannLHC2}
      C.~Kilica and B.~Tweedie,
      arXiv:1211.6106 [hep-ph].


\bibitem{ATLAS2012}
    G.~Aad {\it et al.}  [ATLAS Collaboration],
    Phys.\ Lett.\ B {\bf 716} (2012) 1
    [arXiv:1207.7214 [hep-ex]].

\bibitem{CMS2012}
	  S.~Chatrchyan {\it et al.}  [CMS Collaboration],
	  Phys.\ Lett.\ B {\bf 716} (2012) 30
	  [arXiv:1207.7235 [hep-ex]].
	  
\bibitem{ATLAS2012update}
          G.~Aad {\it et al.} [ATLAS Collaboration],
          ATLAS-CONF-2012-170, Dec. 2012.


\bibitem{Arbey2012}
      A.~Arbey, M.~Battaglia, A.~Djouadi and F.~Mahmoudi,
      arXiv:1211.4004 [hep-ph].

\bibitem{Haber1996}
	H.~E.~Haber, R.~Hempfling and A.~H.~Hoang,
	Z.\ Phys.\ C {\bf 75} (1997) 539
	[arXiv:hep-ph/9609331].

\bibitem{Badziak2012}
	M.~Badziak, E.~Dudas, M.~Olechowski and S.~Pokorski,
	JHEP {\bf 1207} (2012) 155
	[arXiv:1205.1675 [hep-ph]].


\bibitem{SPA2005}
  J.~A.~Aguilar-Saavedra, A.~Ali, B.~C.~Allanach, R.~L.~Arnowitt, H.~A.~Baer, J.~A.~Bagger, C.~Balazs and V.~D.~Barger {\it et al.},
  Eur.\ Phys.\ J.\ C {\bf 46} (2006) 43
  [arXiv:hep-ph/0511344].

\bibitem{PDG2012}
  J.~Beringer {\it et al.}  [Particle Data Group Collaboration],
  Phys.\ Rev.\ D {\bf 86} (2012) 010001.

\bibitem{HFAG}
  D.~Asner {\it et al.} [Heavy Flavor Averaging Group Collaboration],
  arXiv:1010.1589 [hep-ex], and online update at {\tt http://www.slac.stanford.edu/xorg/hfag}.

\bibitem{asl}
{\tt https://twiki.cern.ch/twiki/bin/view/}\\
{\tt AtlasPublic/CombinedSummaryPlots}

\bibitem{csl}
{\tt https://twiki.cern.ch/twiki/bin/view/}\\ 
{\tt CMSPublic/PhysicsResultsSUS}


\bibitem{Passarino:1978jh}
  G.~Passarino and M.~J.~G.~Veltman,
  Nucl.\ Phys.\ B {\bf 160} (1979) 151.

\bibitem{Denner:2010tr}
  A.~Denner and S.~Dittmaier,
  Nucl.\ Phys.\ B {\bf 844} (2011) 199
  [arXiv:1005.2076 [hep-ph]].

\bibitem{Dittmaier:2003bc}
  S.~Dittmaier,
  Nucl.\ Phys.\ B {\bf 675} (2003) 447
  [arXiv:hep-ph/0308246].


\bibitem{FeynArts}
      T.~Hahn,
      Comput.\ Phys.\ Commun.\ {\bf 140} (2001) 418
      [arXiv:hep-ph/0012260].  

\bibitem{FeynCalc}
      R.~Mertig, M.~B\"ohm and A.~Denner,
      Comput.\ Phys.\ Commun.\ {\bf 64} (1991) 345.

\bibitem{FORM}
      J.~A.~M.~Vermaseren,
      arXiv:math-ph/0010025.

\bibitem{DMNLO}
  {\tt http://dmnlo.hepforge.org}.


\bibitem{Baro:2009gn}
  N.~Baro and F.~Boudjema,
  Phys.\ Rev.\ D {\bf 80} (2009) 076010
  [arXiv:0906.1665 [hep-ph]].

\bibitem{Heinemeyer:2010mm}
  S.~Heinemeyer, H.~Rzehak and C.~Schappacher,
  Phys.\ Rev.\ D {\bf 82} (2010) 075010
  [arXiv:1007.0689 [hep-ph]].

\bibitem{Kovarik2005}
  K.~Kova\v{r}\'{\i}k, C.~Weber, H.~Eberl and W.~Majerotto,
  Phys.\ Rev.\  D {\bf 72} (2005) 053010   
  [arXiv:hep-ph/0506021].

\bibitem{MBmass} 
  K.~Melnikov and A.~Yelkhovsky, Phys.\ Rev.\ {\bf D59} (1999) 114009;\\
  A.~H.~Hoang, Phys.\ Rev.\ {\bf D61} (2000) 034005;\\ 
  M.~Beneke and A.~Signer, Phys.\ Lett.\ {\bf B471} (1999) 233;\\
  A.~A.~Penin and A.~A.~Pivovarov, Nucl.\ Phys. {\bf B549} (1999) 217.

\bibitem{Baer2002}
  H.~Baer, J.~Ferrandis, K.~Melnikov and X.~Tata,
  Phys.\ Rev.\  D {\bf 66} (2002) 074007
  [arXiv:hep-ph/0207126].  

\bibitem{QCDhiggs}
  K.~G.~Chetyrkin, Phys.\ Lett.\ B {\bf 390} (1997) 309 [hep-ph/9608318];\\
  P.~A.~Baikov, K.~G.~Chetyrkin and J.~H.~Kuhn, Phys.\ Rev.\ Lett.\  {\bf 96} (2006) 012003 [hep-ph/0511063];\\
  K.~G.~Chetyrkin and A.~Kwiatkowski, Nucl.\ Phys.\ B {\bf 461} (1996) 3 [arXiv:hep-ph/9505358].

\bibitem{Carena2000}
  M.~S.~Carena, D.~Garcia, U.~Nierste and C.~E.~M.~Wagner,
  Nucl.\ Phys.\  B {\bf 577} (2000) 88
  [arXiv:hep-ph/9912516].

\bibitem{Spira2003}
  J.~Guasch, P.~H\"afliger and M.~Spira,
  Phys.\ Rev.\  D {\bf 68} (2003) 115001
  [arXiv:hep-ph/0305101].

\bibitem{Eberl:1999he}
  H.~Eberl, K.~Hidaka, S.~Kraml, W.~Majerotto and Y.~Yamada,
  Phys.\ Rev.\ D {\bf 62} (2000) 055006
  [arXiv:hep-ph/9912463].

\bibitem{Weber:2007id}
  C.~Weber, K.~Kovarik, H.~Eberl and W.~Majerotto,
  Nucl.\ Phys.\ B {\bf 776} (2007) 138
  [arXiv:hep-ph/0701134].


\bibitem{GieleGlover}
      W.~T.~Giele and E.~W.~N.~Glover,
      Phys.\ Rev.\ D {\bf 46} (1992) 1980.

\bibitem{HarrisOwens}
  B.~W.~Harris and J.~F.~Owens,
  Phys.\ Rev.\ D {\bf 65} (2002) 094032
  [arXiv:hep-ph/0102128].

\bibitem{Denner:1991kt}
  A.~Denner,
  Fortsch.\ Phys.\  {\bf 41} (1993) 307
  [arXiv:0709.1075 [hep-ph]].

\bibitem{Catani-Seymour}
      S.~Catani, S.~Dittmaier, M.~H.~Seymour and Z.~Trocsanyi,
      Nucl.\ Phys.\ B {\bf 627} (2002) 189
      [arXiv:hep-ph/0201036].

\bibitem{tHooft:1978xw}
  G.~'t Hooft and M.~J.~G.~Veltman,
  Nucl.\ Phys.\ B {\bf 153} (1979) 365.

\bibitem{Beenakker1997}
      W.~Beenakker, R.~H\"opker, M.~Spira and P.~M.~Zerwas,
      Nucl.\ Phys.\ B {\bf 492} (1997) 51
      [arXiv:hep-ph/9610490].

\bibitem{Tait:1999cf}
  T.~M.~P.~Tait,
  Phys.\ Rev.\ D {\bf 61} (2000) 034001
  [arXiv:hep-ph/9909352].
	
\bibitem{GoncalvesNetto:2012yt}
  D.~Goncalves-Netto, D.~Lopez-Val, K.~Mawatari, T.~Plehn and I.~Wigmore,
  arXiv:1211.0286 [hep-ph].



\bibitem{CalcHEP}
      A.~Pukhov,
      arXiv:hep-ph/0412191.


\end{thebibliography}

\end{document}